\RequirePackage{fix-cm}
\documentclass[twocolumn]{svjour3}          
\smartqed  
\usepackage{graphicx}
\usepackage{url,bm,cite,multirow}

\usepackage[fleqn]{amsmath}
\usepackage{amsfonts}
\usepackage{amssymb}
\DeclareMathOperator*{\argmax}{argmax}

\usepackage{algorithm}
\usepackage{algorithmic}

\journalname{Multimedia Systems}

\begin{document}

\title{Deep learning-based automatic downbeat tracking: a brief review
}

\titlerunning{Deep learning-based automatic downbeat tracking...}        

\author{Bijue Jia*         \and
        Jiancheng Lv*         \and
        Dayiheng Liu* 
}

\institute{Jiancheng Lv \at
              \email{lvjiancheng@scu.edu.cn}           
           \and
           Bijue Jia \at
              \email{jiabijue@outlook.com}
           \and
           Dayiheng Liu \at
              \email{losinuris@gmail.com}
           \and
           * College of Computer Science, Sichuan University, Chengdu, People's Republic of China
}

\date{Received: date / Accepted: date}

\maketitle

\begin{abstract}
As an important format of multimedia, music has filled almost everyone's life. Automatic analyzing music is a significant step to satisfy people's need for music retrieval and music recommendation in an effortless way. Thereinto, downbeat tracking has been a fundamental and continuous problem in Music Information Retrieval (MIR) area. Despite significant research efforts, downbeat tracking still remains a challenge. Previous researches either focus on feature engineering (extracting certain features by signal processing, which are semi-automatic solutions); or have some limitations: they can only model music audio recordings within limited time signatures and tempo ranges. Recently, deep learning has surpassed traditional machine learning methods and has become the primary algorithm in feature learning; the combination of traditional and deep learning methods also has made better performance. In this paper, we begin with a background introduction of downbeat tracking problem. Then, we give detailed discussions of the following topics: system architecture, feature extraction, deep neural network algorithms, datasets, and evaluation strategy. In addition, we take a look at the results from the annual benchmark evaluation--Music Information Retrieval Evaluation eXchange (MIREX), as well as the developments in software implementations. Although much has been achieved in the area of automatic downbeat tracking, some problems still remain. We point out these problems and conclude with possible directions and challenges for future research.

\keywords{Music downbeat tracking\and Music Information Retrieval \and Deep learning \and Multimedia \and Review}
\end{abstract}

\section{Introduction}\label{sec:1_intro}
Music is explicitly structured in a temporal manner. The time structure of a music piece is often conceived as a superposition of multiple hierarchical levels or time-scales~\cite{lerdahl1985generative}. People can synchronize with these temporal scales while playing instruments or dancing. The mensural level of these temporal structures (which people tap their feet to) contains the approximately equally spaced \emph{beat}, which is the basic unit of time and pulse (regularly repeating event) in music theory. Another highly-related term is \emph{tatum}, which is the lowest regular pulse train that a listener intuitively infers from the timing of perceived musical events (i.e. a time quantum). According to music's metrical structure, the same amount of beats are segmented sequentially into groups called \emph{bars} or \emph{measures}. The first beat of each bar plays a role of accentuation, and it is defined as a \emph{downbeat}. Downbeats are often used by composers and conductors to help musicians read and navigate in a musical piece and by music fans and amateurs to better learn music. Automatically analyzing and estimating downbeat is of significant importance when we are trying to analyze and follow a music piece.

\begin{figure*}
  \includegraphics[width=\textwidth]{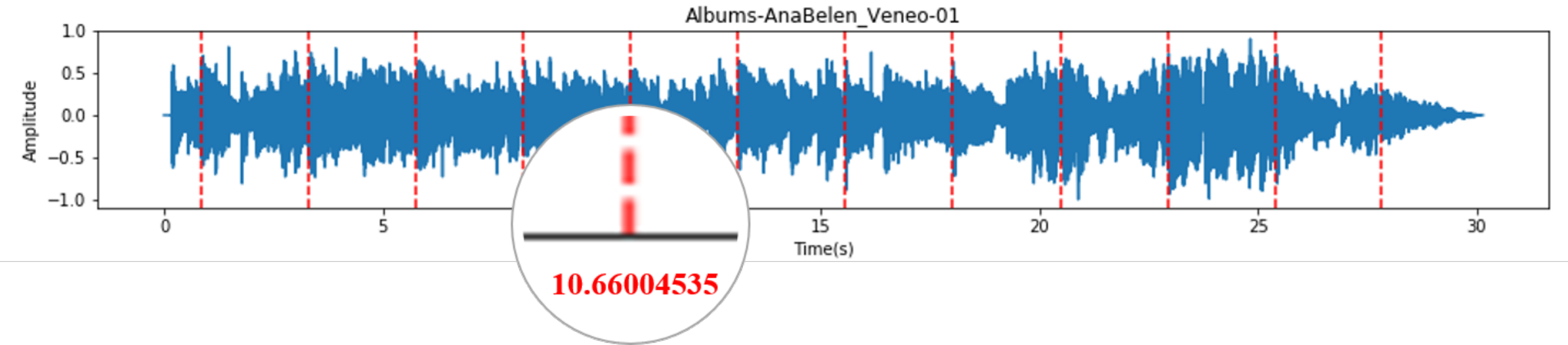}
\caption{Example of a typical downbeat annotation (\emph{Albums-AnaBelen\_Veneo-01.beat} from \emph{Ballroom} dataset), showing downbeat time (in red dashed line).}
\label{fig:annotation_example}
\end{figure*}

The research area that investigates computation models for tracking downbeats is known as \emph{Automatic downbeat tracking} (also called downbeat detection or downbeat estimation). The goal of downbeat tracking is to automatically annotate the time points of all the downbeats in a piece of music audio. An example of a song's annotation file is shown in Fig. \ref{fig:annotation_example}. It is useful for various tasks such as music audio transcription~\cite{sigtia2014rnn, sigtia2016end, sturm2016music, cogliati2016context}, chord recognition~\cite{oudre2011probabilistic, di2013automatic}, structure segmentation~\cite{maddage2006automatic, serra2014unsupervised, panagakis2014elastic, pauwels2013combining, muller2015fundamentals} and musicology analysis. Automatic downbeat tracking can also be used in music information retrieval~\cite{mirex2018url, downie2003music} and music recommendation~\cite{oscar2010music,park2015a,yang2018review}. This problem has long been paid attention to in the community of \emph{Music Information Retrieval} (MIR, which is an interdisciplinary research field focusing on searching and obtaining information from music. Related background knowledge include, but not limited to, musicology, psychology, signal processing, informatics, statistical learning and machine learning\footnote{For a more comprehensive survey of MIR, containing background, history, fundamentals, tasks and applications, we refer readers to the overview by \cite{downie2003music, typke2005survey, casey2008content, downie2008music}.}.) Automatic downbeat tracking has also attracted worldwide scholars to exert their efforts to and has been one of the challenging tasks of Music Information Retrieval Evaluation eXchange (MIREX)~\cite{mirex2018url} in recent years. The level of current interest by its recent inclusion in automatic downbeat tracking problem amongst the community is illustrated and compared in the MIREX evaluation initiative. The most similar task related to downbeat tracking is beat tracking~\cite{goto1994beat, goto1995real, goto2001audio, davies2005beat, seppanen2006joint, gkiokas2012music}, which has been studied much longer than downbeat tracking. A few researchers are also studying these two tasks together~ \cite{peeters2011simultaneous,krebs2013rhythmic,krebs2014unsupervised,krebs2013rhythmic,bock2016joint}. Tracking beats is difficult, however tracking downbeats is comparably hard.

Downbeat tracking problem has been studied from very earlier. The premier one proposed by \cite{goto1999real} models three metrical levels and is reported to successfully track downbeats in 4/4 music with drums, however, it is built upon hand-designed features and patterns. Unfortunately, annotating downbeat positions manually is a time-consuming and expensive process and heavily depends on the intuition of the developer. Hand-crafted features and rules are also not readily available for most music recordings~\cite{davies2006spectral,durand2014enhancing}. A general trend is to divert the attention to automatical methods. Later systems start to go from hand-crafted features to automatically learned ones. One line uses probabilistic state-space models, where rhythmic patterns are learned from data and used as an observation model~\cite{klapuri2006analysis,peeters2011simultaneous,krebs2013rhythmic}. Another line uses Support Vector Machines (SVMs) to track downbeat in a semi-automatic setting~\cite{jehan2005downbeat}, and later transforms into a fully automatic system with a few beat-synchronous hand-annotated features. The system of \cite{davies2006spectral} tracks beats first and then calculates the Kullback-Leibler divergence between two consecutive band-limited beat synchronous spectral difference frames to track downbeats. Papadopoulos and Peeters~\cite{papadopoulos2011joint} jointly tracks chords and downbeats by decoding a sequence of beat-synchronous chroma vectors using Hidden Markov Model (HMM). There are some problems exist in these systems as well: they are applicable to only several simple metrical structures~\cite{klapuri2006analysis,gartner2014unsupervised}, or limited musical styles~\cite{hockman2012one,krebs2013rhythmic,srinivasamurthy2014search}, or restrictive prior knowledge~\cite{allan2004bar,davies2006spectral,papadopoulos2011joint}. Systems that forecast some necessary information beforehand are naturally prone to error propagation.

Recent studies resort to deep learning to try to solve the above problems. As the amount and variousness of data increases, designing features and rules manually is infeasible. Deep learning can obtain higher-level and abstract musical representations that fully characterize the complexity of the problem that is hard to design by hand. Many of these factors of variation can be identified only through sophisticated, nearly human-level understanding of music. Deep learning solves this problem by introducing representations that are expressed in terms of other, simpler representations~\cite{goodfellow2016deep}. The advent of deep learning has had a significant impact on many areas in machine learning and information retrieval, dramatically improving the state-of-the-art in tasks such as object detection, image classification, speech recognition, and language translation. Recent years have also witnessed a deluge of researches in multimedia processing by using deep learning, such as music recommendation, multimedia labeling and retrieval~\cite{wang2014improving, yan2015deep, zou2016deep, nie2017convolutional}. As an important and valuable type of multimedia, music can also be well analyzed by deep learning. The quintessential models of deep learning are multifarious deep neural networks (DNNs). This survey focuses on DNN-based music downbeat tracking, which has achieved intriguing and effective results~ \cite{durand2015downbeat, durand2016feature, durand2017robust, bock2016joint, krebs2016downbeat}.

\begin{figure}
 \centerline{
 \includegraphics[width=0.8\columnwidth]{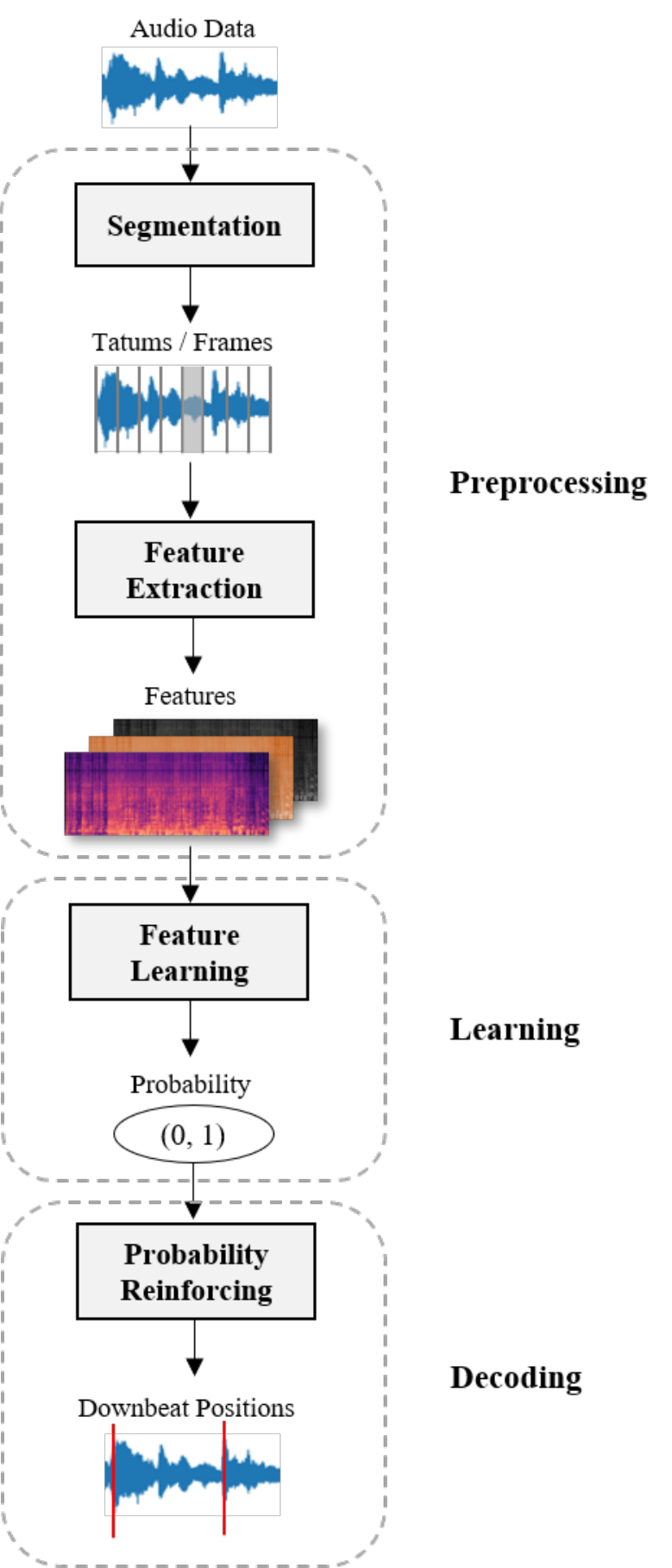}}
 \caption{General architecture for downbeat tracking systems.}
 \label{fig:sys_arch}
\end{figure}

Downbeat tracking problem is a bit similar to classification or sequence labeling problem~\cite{graves2012supervised}, whose aim is to annotate a tag to each segment of the original audio sequence. From an overall perspective, a typical DNN-based automatic downbeat tracking system comprises three major phrases: data preprocessing, feature learning and temporal decoding. An ensemble paradigm of the downbeat tracking system is shown in Fig. \ref{fig:sys_arch}. More particularly, data preprocessing can be separated into two procedures called segmentation and feature extraction, and after feature learning, there is a small procedure called feature combination. Step by step, all of the procedures are:

\textbf{Segmentation}: In all downbeat tracking systems, segmentation would be the first step. By doing so makes it much easier for subsequent stages to detect downbeats because it does not have to deal with tempo or expressive timing on one hand and it greatly reduces the computational complexity by both reducing the sequence length of an excerpt and the search space. Downbeat tracking is then reduced to a classification or sequence labeling problem where each segment is decided as a downbeat or not.

\textbf{Feature Extraction}: After segmenting, every piece of the fragment is a possible candidate of a downbeat. The first thing to do, is amplifying and extracting some signal features so that the latter learning algorithm could capture the characteristic easily. In western music, the downbeats usually coincide with chord changes or harmonic cues, whereas in non-western music the start of a measure is often defined by the boundaries of rhythmic patterns. Therefore, many algorithms exploit one or more of these features to track the downbeats \cite{bock2016joint}. The most likely attributes--which are decided manually using domain-specific knowledge of music--that contribute to the perception of downbeats are harmony, timbre, bass content, rhythmic pattern, the local similarity in timbre and harmony and percussion. Among them, six attributes (harmony, timbre, bass content, rhythmic pattern, the local similarity in timbre and harmony) contribute to the grouping of beats into a bar; two attributes (harmony and percussion) are beat-synchronous features.

\textbf{Feature Learning}: The extracted features are then running through the DNN. If features are more than one kind, each of them is sent to independent neural networks as input, and these networks are called feature adapted neural networks. This is a convenient approach to work with features of different dimension and assess the effect of each of them. More detailed illustration about DNN-based feature learning methods will be discussed in Section 3.

\textbf{Feature Combination}: As it is important for the following decoding process to reduce estimation errors, leading to a tradeoff among the outputs from different feature adapted neural networks. Durand et al.~\cite{durand2015downbeat} use an average of the observation probabilities obtained by those independent networks. The average or sum rule is in general quite resilient to estimation errors~\cite{kittler1998combining}.

\textbf{Temporal Decoding}: Temporal decoding stage is the last step of a downbeat tracking system; it analyzes a downbeat likelihood sequence which is output by DNNs and maps the sequence into a discrete sequence of downbeats. Commonly-used methods are Hidden Markov Model (HMM) and Dynamic Bayesian Network (DBN). Krebs et al.~\cite{krebs2016downbeat} have experimented and proved that an added DBN stage is performing better than a simple DNN output (i.e. simply reports downbeats if the output likelihood of DNN activations exceeds a threshold). In Section 4 we will give a detailed description of each algorithm.

Consequently, to give researchers a clear understanding of DNN-based automatic downbeat tracking system, this review expatiates each step of the system as comprehensive as possible. More importantly, this paper focuses on three different general DNN architectures during the feature learning step and makes a brief comparison among them. Additionally, we further go through some work and information that are involved in downbeat tracking researches.

The remainder of this paper is organized as follows. Section \ref{sec:2_segmentation} gives an overview of the segments and segmentation methods for preprocessing.  In Section \ref{sec:3_feature_extraction}, we will summarize all the features that correlate to downbeats and their extraction methods. Common and general DNN models are depicted in Section \ref{sec:4_dnns}. In the following Section \ref{sec:5_temporal_decoding}, several frequently used temporal decoding and machine learning algorithms will be summed-up. Section \ref{sec:6_datasets} gives a complete list of datasets used for downbeat tracking problem. Some commonly-used evaluation methods are discussed in Section \ref{sec:7_evaluation}. Then in Section \ref{sec:8_software} describes an incomplete list of the most relevant software packages or libraries to downbeat tracking. Finally, Section \ref{sec:9_discussion} discusses the prevalent methods and what the probable future directions are and what the most challenging issues could be. This survey is structured logically rather than chronologically.

\section{Segments and Segmentation Methods}\label{sec:2_segmentation}
The goal of music audio segmentation is to switch downbeat annotation problem to sequence labeling problem. Finding the exact timestamp of a downbeat is impossible because time is continuous. Instead, we can split music audio into a sequence of small segments and decide each segment is a downbeat or not. If a segment is a downbeat, then use the occurrence time of this segment as the annotation of this downbeat. Doing segmentation is desirable because tempo-invariant features decrease the capacity and simplify the feature learning process while making it less prone to over-fitting~\cite{durand2015downbeat}. There are three kinds of segments that are commonly used in downbeat tracking: beat segment, tatum segment, and frame segment.

\textbf{Beat Segmentation}: Durand et al.~\cite{durand2015downbeat} and Krebs et al.~\cite{krebs2016downbeat} temporally segment the signal into subdivisions of the rhythmic beat. They seek the segmentation that maximizes downbeat recall rate while emphasizing consistency in inter-segment durations. To achieve these goals they extend the \emph{local pulse information extractor} presented in \cite{grosche2011tempogram} and process the following operations: a) First, they use this toolbox to obtain a tempogram of the musical audio. b) Then they use dynamic programming with strong continuity constraints and emphasis towards high tempi. c) Finally they use the decoded path to recover instantaneous phase and amplitude values, construct the predominant local pulse (PLP) function as in \cite{grosche2011tempogram}, and detect pulses using peak-picking~ \cite{dixon2007evaluation}. Using this procedure, the recall rate for downbeat pulses is above 95\% for each dataset, using a 100 ms tolerance window.

\textbf{Tatum Segmentation}: Durand et al.~\cite{durand2016feature,durand2017robust} adopt the \emph{local pulse information extractor} proposed to achieve a useful tatum segmentation. The Processing procedure is: a) Computing the tempogram of the musical audio through a Short-Term Fourier Transform (STFT) and only keep the tempo above 60 repetitions per minute to avoid slow metrical levels. b) Tracking the best periodicity path by dynamic programming with the same kind of local constraints. The following system can find a fast subdivision
of the downbeats at a rate that is locally regular. c) Finally using the decoded path to recover instantaneous phase and amplitude values, construct the PLP function, and detect tatums using peak-picking on the PLP. The resulting segmentation period is typically twice as fast as the beats period, while it can be up to four times faster.

\textbf{Frame Segmentation}: B\"ock et al.~\cite{bock2016joint} use a very simple way to do segmentation. They split audio into overlapping frames, with 100 frames per second (100 fps), implying that two neighboring frames are located 10ms apart. This is also the initial processing stage of STFT. Unlike beat and tatum, a frame is not as a low-level music characteristic but as a raw audio piece. Using frame segmentation avoids hand-crafted features such as harmonic change detection~ \cite{durand2014enhancing,durand2015downbeat,durand2016feature,khadkevich2012probabilistic,papadopoulos2011joint}, or rhythmic patterns~ \cite{hockman2012one,holzapfel2014tracking,krebs2013rhythmic}. The relevant features can be learned directly from the spectrogram, therefore frame segmentation shows up in pair with auto-learned features, and should co-operate with DNN-based feature learning algorithms (will be discussed in section \ref{subsec:3.9auto-learned_feature} and section \ref{sec:4_dnns}).

\section{Features and Feature Extraction Algorithms}\label{sec:3_feature_extraction}
Finding musical features that correlate to downbeat is very helpful since these attributes make learning algorithms or classifiers to apperceive downbeats more easily. It is worth mentioning that in most cases hand-crafted feature works well when the dataset is not large, homogenous, high-qualified or identically-distributed. By doing features extraction, the dimension of data is reduced, so that learning algorithms could be less complicated and run faster. In this section, we summarize the most relevant features to downbeats and their corresponding extraction methods.

\subsection{Harmony}\label{subsec:3.1harmony}
In music, harmony considers the process by which the composition of individual sounds, or superpositions of sounds, is analyzed by hearing. Usually, this means simultaneously occurring frequencies, pitches (tones, notes), or chords~\cite{malm1996music}. Change in harmony or timbre content (will be described in section \ref{subsec:3.3timbre}), for example, chord changes, section changes or the entrance of a new instrument is often related to a downbeat position~\cite{durand2015downbeat}. The feature of harmony is represented by chroma~\cite{bello2005robust}.

There are two main ways of extracting harmony. One way is used by Durand et al.~\cite{durand2015downbeat,durand2017robust} and they extracts the harmonic feature as the following steps: 1)First, they down-sample the audio signal at 5512.5 Hz. 2)They then compute the STFT using a Hann window of size 4096 and a hop size of 512. 3)They apply a constant-Q filter-bank with 108 bins (36 bins per octave). 4)They convert constant-Q spectrum to harmonic pitch class profiles 5)Afterward, they remove octave information by accumulating the energy of equal pitch classes. 6)They tune the chromagram by finding bias on peak locations; smooth it by a median filter of length 8. 7)In the end, they map it to a 12 bins representation by averaging. The other way is conducted by Krebs et al.~\cite{krebs2016downbeat} and they use the CLP chroma feature~\cite{muller2011chroma} with a frame rate of 100 frames per second. Then they synchronize the features to the beat by computing the mean value over a window of length ${\triangle{b}} / n_h$ (${\triangle{b}}$ is the beat period), yielding $n_h=2$ feature values per beat interval.

\subsection{Harmony Similarity}\label{subsec:3.2harmony_sim}
By looking at harmony similarity or timbre similarity (detailed description is in section \ref{subsec:3.4timbre_sim}), we can observe longer-term patterns of change and novelty that are invariant to the specific set of pitch values or spectral shape. The similarity in harmony, for example, has the interesting property of being key invariant and therefore can model cadences and other harmonic patterns related to downbeat positions~ \cite{durand2015downbeat}. The feature of harmony similarity is represented by chroma similarity (CS).

The chromas are computed the same as in section \ref{subsec:3.1harmony}, but they are then averaged to obtain segment synchronous chroma. For each segment, compute the cosine similarity of one segment synchronous chroma with the 24 segment synchronous chroma around it. The dimension of CS is 24.

\subsection{Timbre}\label{subsec:3.3timbre}
In music, timbre is the perceived sound quality of a musical note, sound or tone. Timbre distinguishes different types of sound production, such as choir voices and musical instruments. Alternations of the timbre-inspired content occur more likely at the start of a new section and near a downbeat position~\cite{durand2017robust}. the feature of timbre is represented by Mel-frequency cepstral coefficients (MFCC). Timbre extraction can also be done in conjunction with an onset~\cite{jehan2005downbeat}, tatum or beat segmentation~\cite{durand2017robust}.

Durand et al.~\cite{durand2015downbeat} compute the first 12 Mel-frequency cepstral coefficients using \cite{voicebox}, with a Hamming window of size 2048, a hop size of 1024 and 32 Mel filters on a signal sampled at 44100 Hz.

\subsection{Timbre Similarity}\label{subsec:3.4timbre_sim}
The feature of timbre similarity is represented by MFCC similarity (MS). The MFCC spectrograms are computed the same as in section \ref{subsec:3.3timbre}, but they are then averaged to obtain segment synchronous MFCC spectrogram. For each segment, computing the cosine similarity of one segment-synchronous MFCC spectrogram with the 24 segment-synchronous MFCC spectrogram around it. The dimension of MS is 24.

\subsection{Bass Content}\label{subsec:3.5bass_content}
The bass content is low-frequency, containing mostly bass instrument or kick drum, both of which tend to be used to emphasize the downbeat~ \cite{durand2015downbeat}. The feature of low-frequency content is represented by low-frequency spectrogram (LFS).

Durand et al.~\cite{durand2015downbeat,durand2017robust} compute LFS as follows: 1)First they downsample audio signal at 500 Hz. 2)Then they compute STFT by a Hann window of size 32 and a hop size of 4 to get the spectrogram. 3)They keep the spectral components below 150 Hz (the first 10 bins). 4)Finally, they clip the signal so that all values on the 9th decile are equal.

\subsection{Rhythmic Pattern}\label{subsec:3.6rhythm}
Rhythm is the timing of musical sounds and silences that occur over time. Rhythmic patterns are frequently repeated each bar and are therefore useful to obtain the bar boundaries. The feature of the rhythmic pattern is represented by onset detection function (ODF).

Durand et al. in their early work~\cite{durand2015downbeat} use 4 band-wise ODF as computed by \cite{klapuri2006analysis}. First, they compute the STFT using a Hann window of size 1024 and a hop size of 256 for a signal sampled at 44100 Hz. Second, they compute the spectrogram and apply a 36-bands Bark filter. Third, they use $\mu$-law compression with $\mu=100$ and downsample the signal by a factor of two. Fourth, they do envelope detection using an order 6 Butterworth filter with a 10 Hz cutoff. Fifth, a weighted sum of 20\% of the envelope and 80\% of its difference is done to compute the ODF. Finally, they map ODF to 4 equally distributed bands.

Durand et al. in their later work~\cite{durand2017robust,durand2016feature} compute a 3-band spectral flux ODF: 1)They perform STFT to get the spectrogram. 2)They apply $\mu$-law compression with $\mu=10_6$ to the STFT coefficients. 3)They sum the discrete temporal difference of the compressed signal on 3 bands for each temporal interval, and subtract the local mean and half wave. The frequency intervals of the low, medium and high-frequency bands are [0 150], [150 500] and [500 11025] Hz respectively.

\subsection{Melody}\label{subsec:3.7melody}
A melody is a linear succession of musical tones that the listener perceives as a single entity; it is a combination of pitch and rhythm. For melody, some notes tend to be more accented than others and both pitch contour and note duration play important roles in our interpretation of meter~\cite{pfordresher2003role,hannon2004role,ellis2009role}. The feature of melody is represented by melodic constant-Q transform (MCQT).

Durand et al.~\cite{durand2016feature,durand2017robust} get melody features as follows: 1)They downsample audio at 11025 Hz. 2)They conduct STFT with Hann window of size 185.8 ms and hop size 11.6 ms. 3)They apply a constant-Q transform (CQT) with 96 bins per octave, starting from 196 Hz to the Nyquist frequency to the STFT, and average the energy of each CQT bin q[k] with the following octaves:
\begin{equation}
s[k]=\frac {\sum _{j=0}^{J_k}{q[k+96j]}}{J_k+1}
\end{equation}
with $J_k$ such that $q[k + 96Jk]$ is below the Nyquist frequency. 4)Then they only keep 304 bins from 392 Hz to 3520 Hz that correspond to 3 octaves and 2 semitones. 5)They use a logarithmic representation of $s$ to represent the variation of the energy more clearly:
\begin{equation}
r=log(\left| { \hat {s}  } \right| +1)
\end{equation}
where ${ \hat {s} }$ is the restriction of $s$ between 392 Hz and 3520 Hz. 6)They set every value which is under the 3rd quartile $Q_3$ of a given temporal frame to zero to get the final melodic CQT:
\begin{equation}
{ m }_{ CQT }=max(r-{ Q }_{ 3 }(r),\quad 0)
\end{equation}

\subsection{Percussion}\label{subsec:3.8percussion}
Percussion is commonly referred to as "the backbone" or "the heartbeat" of a musical ensemble, often working in close collaboration with bass instruments, when present.

Krebs et al.~\cite{krebs2016downbeat} compute a multi-band spectral flux: 1)First, they compute the magnitude spectrogram by applying the STFT with a Hann window, hop size of 10ms, and a frame length of 2048 samples. 2)Second, they apply a logarithmic filter bank with 6 bands per octave, covering the frequency range from 30 to 17 000 Hz, resulting in 45 bins in total. 3)Third, they compress the magnitude by applying the logarithm. 4)For every frame, they compute the difference between the current and the previous frame. 5)Finally, they beat-synchronize the feature sequence by only keeping the mean value per frequency bin in a window of length ${\triangle{b}} / n_p$, where $\triangle{b}$ is the beat period and $n_p=4$ is the number of beat subdivisions, centered around the beginning of a beat subdivision.

\subsection{Auto-learned Features}\label{subsec:3.9auto-learned_feature}
The selection of appropriate features is a difficult task. Researchers are frequently unsure about which features are useful, and it is difficult to extract the perfect features. Although researchers generally just formulate a limited hypothesis about which type of features may be suitable according to their experience and domain knowledge, this may lead to poor performance because of the limitation of their hypothesis. Spontaneously, we may want machine itself to automatically find out which features are related to downbeat.

A study of automatic extracting features has conducted by \cite{bock2016joint}, who avoids hand-crafted features but prefer the algorithm to learn some relevant features directly from spectrograms. These spectrograms are obtained as follows: 1)Splitting audio signal overlapping frames and weighted with a Hann window of the same length before being transferred to a time-frequency representation with STFT. Two adjacent frames are located 10 ms apart, which corresponds to a rate of 100 fps (frames per second). 2)Omitting the phase portion of the complex spectrogram and use only the magnitudes for further processing. 3)Using three different magnitude spectrograms with STFT lengths of 1024, 2048, and 4096 samples (at a signal sample rate of 44.1 kHz). 4)Limiting the frequencies range to [30, 17000] Hz to reduce the dimensionality of the features. 5)Processing the spectrograms with logarithmically spaced filters. A filter with 12 bands per octave corresponds to semitone resolution, which is desirable if the harmonic content of the spectrogram should be captured. 6)Using filters with 3, 6, and 12 bands per octave for the three spectrograms obtained with 1024, 2028, and 4096 samples, respectively, accounting for a total of 157 bands. 7)Scaling the resulting frequency bands logarithmically to better match human perception of loudness. 8)Adding the first order differences of the spectrograms to the features. The final dimension of the features 314.

\section{DNN-Based Feature Learning Algorithms}\label{sec:4_dnns}
So far, we have introduced downbeat-related music features. The aforementioned features can directly flow into the temporal decoding procedure to get the final results, as some systems do~\cite{MIREX2014KSH1,MIREX2014FK3,MIREX2014FK4,MIREX2015FK,MIREX2016CD4,MIREX2016DSR1}. This works well when data is little, but as the number of data increases, the diversity and complexity of data also grow and some weak points may appear. Under this circumstance, the DNN-based feature learning algorithms are inserted in between the feature extraction and temporal decoding procedures to further extract and learn features. The differences between systems with and without DNN process exist in several aspects:
\begin{itemize}
 \item The fundamental feature extraction algorithms learn more low-level features, while DNN-based feature learning algorithms discover more high-level and abstract features.
 \item The aforementioned features are hand-crafted and empirically-based which heavily resort to prior knowledge of experts and need a very long period to verify effectiveness, while features learned by DNNs are automatically-extracted which rely on the strength of big data and can be verified quickly.
 \item More human prejudices exist in features designed by experts but less in those extracted by learning algorithms. Note that features discovered by learning algorithms may not in sync with our common sense, however they play a vital role in improving model performance.
 \item Using DNN enlarges the number of parameters so that the representation is more powerful.
\end{itemize}
In the following, we will describe and compare three different DNN algorithms. These are the three main models used in feature learning part: Multi-Layer Perceptron (MLP), Convolutional Neural Network (CNN) and Recurrent Neural Network (RNN).

\subsection{Multi-Layer Perceptron}\label{subsec:4.1mlp}

\begin{figure}
 \centerline{
 \includegraphics[width=\columnwidth]{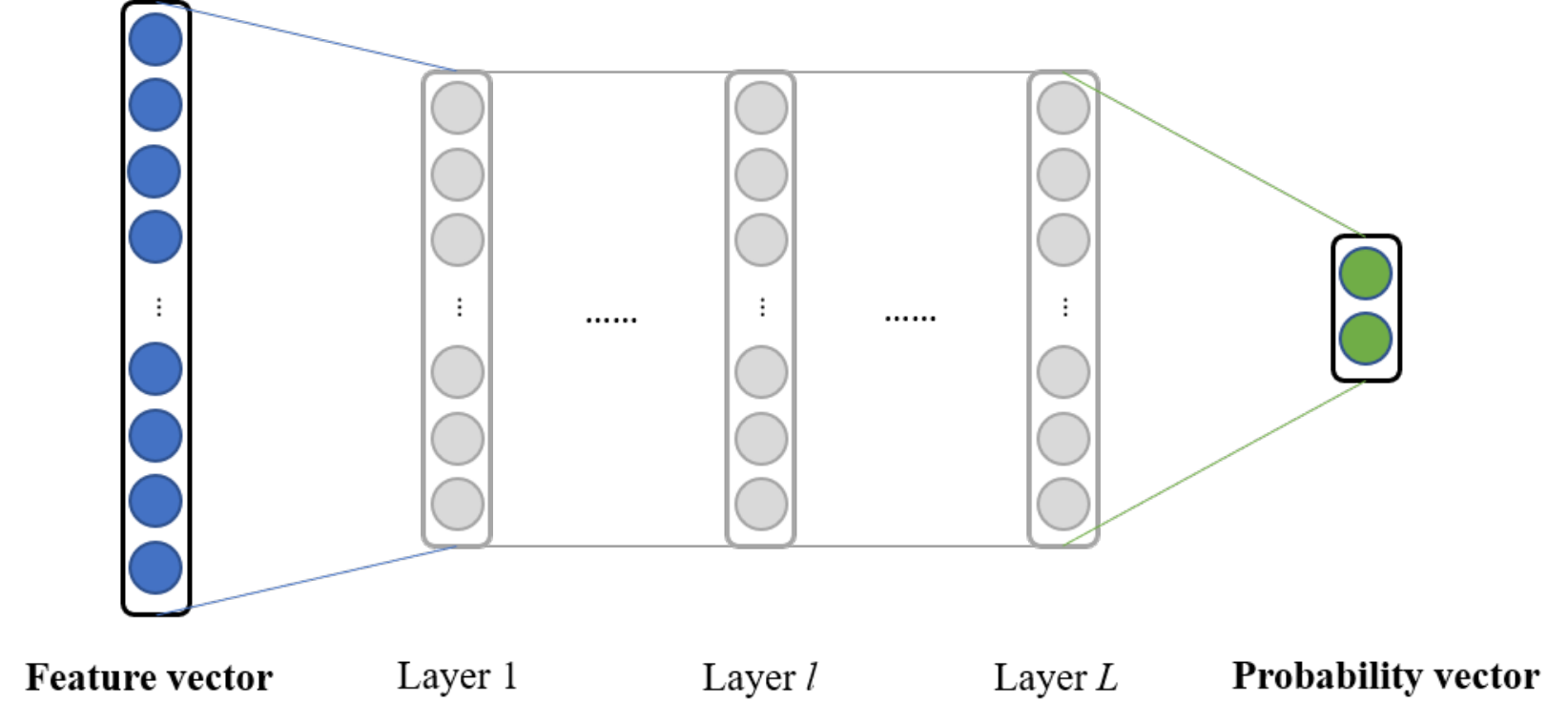}}
 \caption{Sketch diagram of a deep feed forward network.}
 \label{fig:dnn}
\end{figure}

Multi-Layer Perceptron, or deep feedforward network, is the quintessential deep learning model. In some papers, it is also called the general DNN~ \cite{durand2015downbeat}. In order not to cause ambiguity, we refer to MLP instead of DNN when we talk about this algorithm. While playing the role of feature learner in downbeat tracking problems, MLP is a series of functions to estimate the probability of a feature being a downbeat. A sketch diagram of MLP is shown in Fig. \ref{fig:dnn}. An MLP is a cascade of $L$ layer functions of performing linear and non-linear transformations successively. The $l$-th layer functions are:
\begin{equation}
{\mathbf{z}}_l = {{\mathbf{x}_{l-1}}{\mathbf{W}}_{l-1}}+{\mathbf{b}}_{l-1},
\end{equation}
\begin{equation}
f_l({\mathbf{x}}_l; {\boldsymbol{\theta}}_l) = \varphi({\mathbf{z}}_l), \quad \boldsymbol{\theta} = [\mathbf{W}_l; \mathbf{b}_l]
\end{equation}
where $\mathbf{x}_l \in {\mathbb{R}}^{d}$ with dimension $d$ is the input downbeat feature vector when $l=1$, and the output value of layer $l-1$ when $l>1$. $\varphi$ is the non-linear transformation function (e.g. sigmoid, ReLU \cite{glorot2011deep}, maxout \cite{goodfellow2013maxout}, etc.). $\boldsymbol{\theta}_l$ represents the $l$-th layer parameters; $\mathbf{W}_l \in {\mathbb{R}}^{d \times d_{l}}$ is a matrix of weights; $\mathbf{b}_l \in {\mathbb{R}}^{d_{l}}$ is a vector of biases; $d_{l}$ is the dimension of layer $l$. At $L$-th layer (the output layer), $\varphi$ is normally the sigmoid function:
\begin{equation}
sigmoid({\mathbf{z}}_L) = {1 \over {1 + e^{ - {\mathbf{z}}_L}}},
\end{equation}
or the softmax function:
\begin{equation}
softmax{ ({\mathbf{z}}_{L}) }_{i} = \frac { exp{({\mathbf{z}}_{L})}_{i} }{ \sum _{k=1}^{K}{ { exp({\mathbf{z}}_{L})}_{k} }  } .
\end{equation}
where $K$ is the dimensionality of the output layer and also is the number of classes we want to detect. Sigmoid can only be used for binary classification issue, while softmax can deal with more than two classes. Both of them output conditional probabilities ${\rm{P}}({\mathbf{x}}_1|\Theta)$. As for downbeat tracking problem, sigmoid function just gives the probability of one feature $\mathbf{x}_1$ being a downbeat (i.e. downbeat likelihood), while softmax gives every probability of one feature belonging to each class.

\subsection{Convolutional Neural Networks}\label{subsec:4.2cnn}
\begin{figure*}
 \centerline{
 \includegraphics[width=\textwidth]{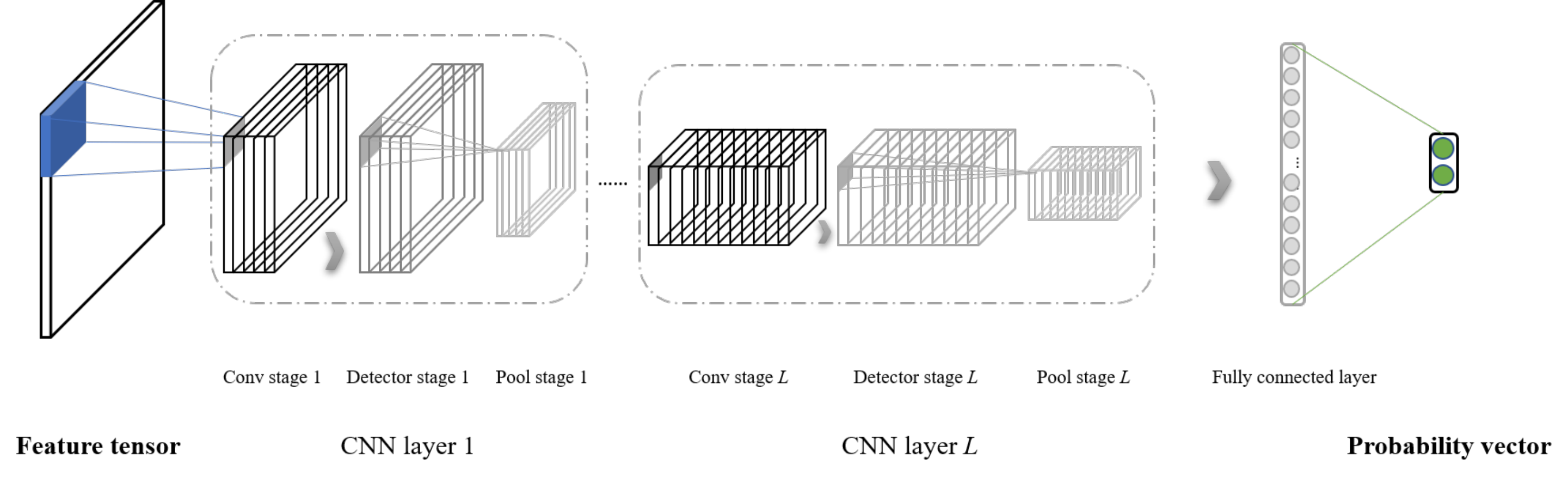}}
 \caption{Sketch diagram of a deep convolutional neural network.}
 \label{fig:cnn}
\end{figure*}

Convolutional Neural Networks are simply neural networks that use convolution in place of general matrix multiplication in at least one layer~ \cite{goodfellow2016deep}. A typical CNN layer consists of three stages sequentially: convolution stage, detector stage, and pooling stage~\cite{goodfellow2016deep}. The complete CNN includes stacked convolutional and pooling layers, at the top of which are multiple fully-connected layers. A sketch diagram of CNN is shown in Fig. \ref{fig:cnn}.

\subsubsection{Convolution Stage}
Given input downbeat features ${\mathbf{X}} \in {\mathbb{R}}^{c \times w \times h}$ with channel number $c$, feature width $w$ (could be time length), and feature height $h$ (could be frequency bandwidth), the convolutional layer convolves $\mathbf{X}$ with $K$ \emph{filters} (or called \emph{kernels}) where each filter ${\mathbf{W}_{k}} \in {\mathbb{R}}^{c \times m \times n}$ is a 3-dimensional tensor with width $m$ and height $n$. We will obtain $K$ feature maps, which constitute a 3-dimensional tensor ${\mathbf{Z}} \in {\mathbb{R}}^{K \times {w}_{\mathbf{Z}} \times {h}_{\mathbf{Z}}}$. The $k$-th feature map ${\mathbf{Z}}_k$ is computed as follows:
\begin{equation}
{\mathbf{Z}}_{k} = {\mathbf{X}} \ast {{\mathbf{W}}_{k}} + {b_{k}}, \quad k=1, \cdots, K.
\end{equation}
where $\ast$ denotes the convolution operation and ${b_{k}}$ is a bias parameter.  The convolution on ${\mathbf{X}}$ is operated not only along the feature height (frequency) axis but also along the feature width (time) axis, which results in a simple 2-dimensional convolution commonly used in computer vision.

\subsubsection{Detector Stage}
Before doing the pooling part, we often operate an element-wise non-linear function on the feature maps we obtain after convolution. Here we also denote ${\varphi}$ as the non-linear function, and transform feature maps ${\mathbf{Z}}$ to ${\mathbf{A}}$:
\begin{equation}
{\mathbf{A}} = \varphi ({\mathbf{Z}})
\end{equation}

\subsubsection{Pooling}
After the element-wise non-linearities, feature maps are passed through a pooling layer. A pooling function replaces the neuron values of the feature map at a certain location with a summary statistic of the nearby neuron values.

The most frequently-used pooling function is max pooling. The max pooling~\cite{zhou1988computation} operation reports the maximum output within a rectangular neighborhood. With regard to the $k$-th activated feature map ${{\mathbf{A}}_k}$, the value at position $(t, r)$ of the after-pooling feature map ${{\mathbf{S}}_k}$ is computed by:
\begin{equation}
[{\mathbf{S}}_k]_{t,r} = {max}_{i=1}^{p} \{[{\mathbf{A}}_k]_{t \times s + i, r \times s + i}\}
\end{equation}
where $s$ is the step size and $p$ is the pooling size. Other popular pooling functions include the average of a rectangular neighborhood, the L2 norm of a rectangular neighborhood, or a weighted average based on the distance from the central pixel. We do pooling only along the frequency axis since it helps to reduce spectral variations while pooling in time has been shown to be less helpful~\cite{sainath2013improvements}.

On the top of the complete CNN, fully-connected layers are applied. Their structures are simply the same as the aforementioned MLP. The input to this fully-connected layer is a concatenation of all flattened feature maps ${{\mathbf{S}}_k}$. The output is the downbeat likelihood.

\subsection{Recurrent Neural Networks}\label{subsec:4.3rnn}
Recurrent neural networks or RNNs~\cite{rumelhart1986learning} are a family of neural networks for processing sequential data. Much as a CNN is a neural network that is specialized for processing a grid of values ${\mathbf{X}}$ such as an image, an RNN is a neural network that is specialized for processing a sequence of values ${\mathbf{x}}_{1}, \cdots , {\mathbf{x}}_{T}$.

\begin{figure}
 \centerline{
 \includegraphics[width=\columnwidth]{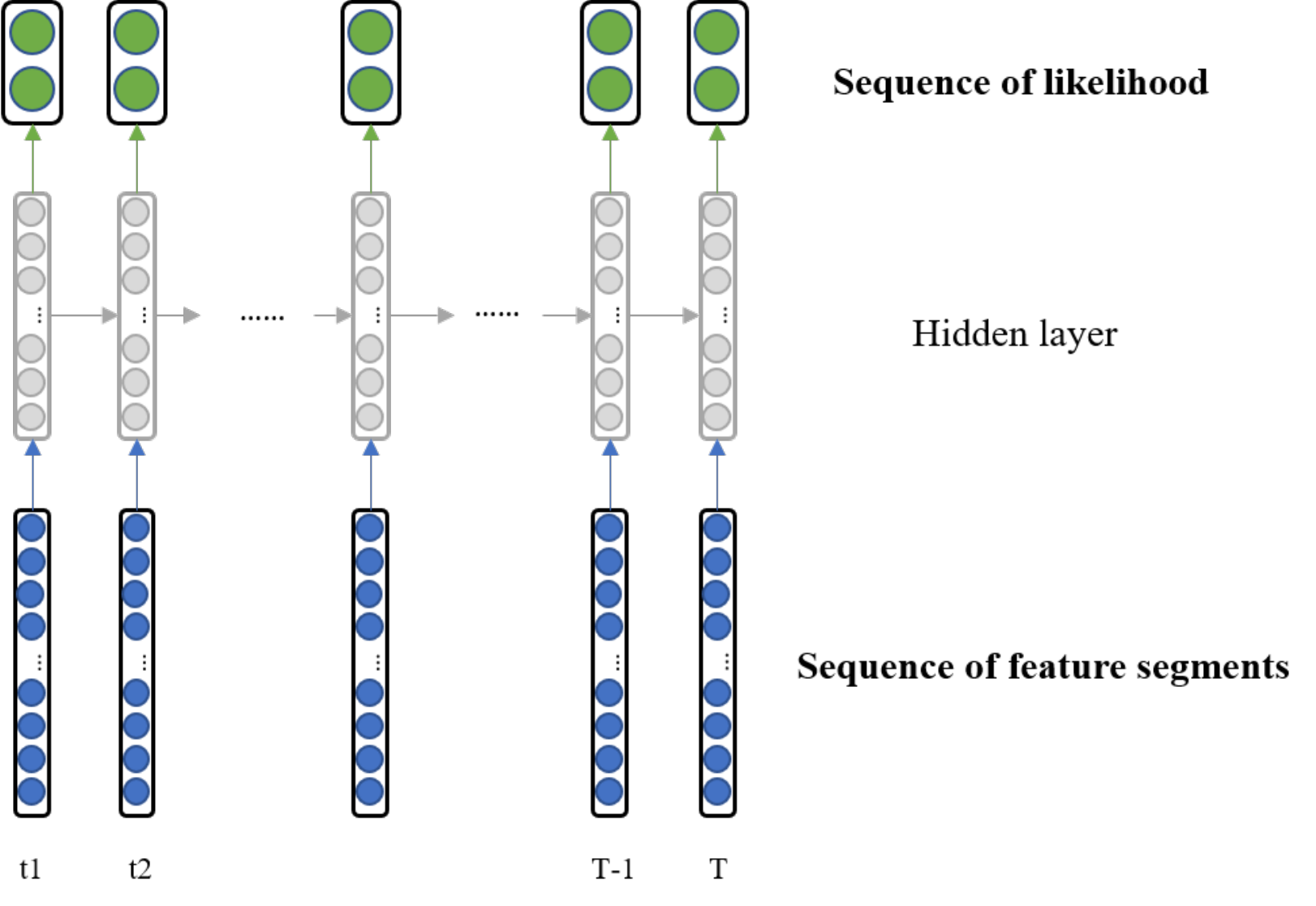}}
 \caption{Sketch diagram of a one-hidden-layer recurrent neural network.}
 \label{fig:rnn}
\end{figure}

Considering the downbeat feature as a sequence ${\mathbf{X}} = [ {\mathbf{x}}_{1}, \cdots, {\mathbf{x}}_{T} ]^{\top}$, in this way, downbeat tracking can be seen as a sequence labeling problem. Vector ${\mathbf{x}}_{t}$ is indexed with time step $t$, ranging from $1$ to $T$. A one-hidden-layer vanilla RNN is composed of three layers: an input layer, a hidden layer, and an output layer. Computation runs along both layer axis and time axis. A one-hidden-layer vanilla RNN is shown in Fig. \ref{fig:rnn}. The hidden layer at time $t$ is computed as:
\begin{equation}
{\mathbf{h}}_{t} = {f_h} ( {{\mathbf{x}}_t} {{\mathbf{W}}_{ih}} + {{\mathbf{h}}_{t-1}} {{\mathbf{W}}_{hh}} + {{\mathbf{b}}_h} )
\label{basic_rnn_unit}
\end{equation}
where ${f_h}$ is the hidden layer activation function, ${{\mathbf{W}}_{ih}}$ is the weight matrix connecting input layer and hidden layer, ${{\mathbf{W}}_{hh}}$ is the weight matrix between adjacent time-step hidden layers (i.e. this weight matrix is shared along the time axis) and ${{\mathbf{b}}_h}$ is the bias vector of the hidden units. Formula \ref{basic_rnn_unit} is also called the basic RNN unit. Output layer at time $t$ is computed as:
\begin{equation}
{\mathbf{y'}}_{t} = {f_o} ( {{\mathbf{h}}_t} {{\mathbf{W}}_{ho}} + {{\mathbf{b}}_o} )
\end{equation}
where ${f_o}$ is the output layer activation function, ${{\mathbf{W}}_{ho}}$ is the weight matrix between hidden layer and output layer and ${{\mathbf{b}}_o}$ is the bias vector of the output units.

Practically, vanilla RNN is not performing well cause its gradient vanishing and exploding issue. More sophisticated and powerful RNN units include Long-Short Term Memory (LSTM)~\cite{hochreiter1997long}, Gated Recurrent Unit (GRU)~\cite{cho2014learning} etc.

If ${f_o}$ is sigmoid function, at time $t$ output $y_t$ is a scalar; the whole output sequence $\mathbf{y'} = [ y'_1, y'_2, \cdots, y'_T ]$ represent the downbeat likelihood. If ${f_o}$ is softmax function and two classes (downbeat and non-downbeat) to be classified, output $\mathbf{y'}_{t} \in {\mathbb{R}}^{2}$ at time $t$ is a vector, consisting of the downbeat likelihood and non-downbeat likelihood value. We take out only the downbeat likelihood values as the final probability sequence $\mathbf{y'} = [y'_1, y'_2, \cdots, y'_T ]$. Each element of this sequence is the parallel-corresponding prediction to the input segment sequence.

\subsection{Comparison of DNNs on Downbeat Tracking}\label{subsec:4.4comparison_of_dnns}
The preceding text has expatiated three kinds of DNN models and described each model independently, however, there are some notable differences among them when solving downbeat tracking problem. The differences are discussed from several perspectives:
\begin{itemize}
 \item From the innate and intrinsic difference point of view, MLP is more computationally expensive due to its fully-connected architecture. Comparing to MLP, the number of parameters of CNN and RNN is much smaller. When the dimension of the downbeat feature is small, these three models all work well. MLP always is the first thought~\cite{durand2015downbeat} because it's very flexible and the results can be used as a baseline point of comparison.
 \item Choosing which model to use also depends on what basic problem the authors see downbeat tracking as. Some researchers view downbeat tracking as a sequence modeling problem. Feature values that fall into one time unit (beat, tatum or frame) are condensed into one vector and all vectors of one audio signal are organized in sequence according to their occurrence time. In this case, RNN is the most suitable model since it is the natural choice for sequence modeling tasks~\cite{krebs2016downbeat,bock2016joint}. CNN can also be used to model sequence and give the probability of each component of a sequence being a downbeat or not, just like the rhythmic neural network designed by \cite{durand2016feature}. Some other researchers treat downbeat tracking as a binary classification problem first--let the model learn to distinguish which input feature is a downbeat feature and which is not. In this case, MLP and CNN are more suitable models~ \cite{durand2015downbeat,durand2016feature,durand2017robust}. When seeing it as binary classification problem, some of the features at non-downbeat position need to be randomly removed in order to obtain an equal amount of features computed at downbeat and non-downbeat positions~\cite{durand2015downbeat}. Each downbeat-correlated musical feature is considered independently and one network is trained per feature. Output probabilities obtained by these independent networks are averaged or summed in the end and then are organized in a probability sequence. Note that when training these classifiers, the temporal correlation between adjacent features is ignored.
\end{itemize}
Generally speaking, when the downbeat problem size is small and we want to quickly get a rough result, MLP would be the first to try; when we want to focus on the spatial relationship within features (such as harmony and melody features), CNN would be better; when we want to model temporal characteristics while learning features, RNN is the better candidate model.

\section{Temporal Decoding Algorithms}\label{sec:5_temporal_decoding}
Temporal decoding maps the output likelihood sequence of DNN into the discrete sequence of downbeats, incorporating musical prior knowledge into the process. Two frequently-used algorithms are HMM and DBN (in fact, HMM is a simple and special case of DBN).  In this section, we will describe in detail the two algorithms and how they solve the last problem.

\subsection{Hidden Markov Model}\label{subsec:5.1HMM}
Hidden Markov Model (HMM) is a probability model with respect to time series. It describes a process, where a hidden Markov chain randomly generates an invisibly random state sequence, and then each state generates each observation. Suppose $S=\{ s_1, s_2, \cdots, s_N \}$ is the set of all possible states, namely the state space; $V = \{ v_1, v_2, \cdots, v_M \}$ is the set of all possible observations. HMM model is composed of three components: initial state probability vector ${\bm{\pi}} \in {\mathbb{R}}^{T}$, state transformation probability matrix ${\mathbf{A}} \in {\mathbb{R}}^{N \times N}$ and observation probability matrix ${\mathbf{B}} \in {\mathbb{R}}^{N \times M}$, where $T$ is the time length. So a HMM model $\lambda$ can be symbolized as:
\begin{equation}
\lambda = ({\mathbf{A}}, {\mathbf{B}}, {\bm{\pi}})
\end{equation}
There are three fundamental problems with regard to HMM: a) probability computation, b) learning problem, and c) decoding problem. Among them, the decoding problem is what we try to solve at the last step of a downbeat tracking system. Decoding problem is defined as this: given model $\lambda=({\mathbf{A}}, {\mathbf{B}}, {\bm{\pi}})$ and observation sequence $\mathbf{o}=[o_1, o_2, \cdots, o_T]$, find the state sequence $\mathbf{y}=[y_1, y_2, \cdots, y_T]$ so that the conditional probability $P(\mathbf{y} | \mathbf{o})$ achieves maximum (i.e. find the most possibly corresponding state sequence).

\subsubsection{Viterbi Algorithm}
Viterbi Algorithm is proposed to solve the decoding problem of HMM by using dynamic programming. It is using dynamic programming to find a path that achieves maximum or best probability; here a path corresponds to a state sequence.

Viterbi algorithm is used to decode downbeat likelihood to the most likely downbeat state sequence~ \cite{durand2017robust,durand2015downbeat,durand2016feature}. They model the problem as follows:

1) State space $S = \{ s_1, s_2, \cdots, s_N \}$, where $N$ is the number of possible states. On the whole, states are partitioned into two distinct states: downbeat and non-downbeat. It is worth noting that the downbeat likelihood depends on the bar length and the position inside a bar, therefore a state is defined for each possible segment (beat or tatum) in a given bar. For those~\cite{durand2015downbeat} who segment audio signal into beat segments, states correspond to downbeats and non-downbeats in a specific metrical position. For example, the downbeat in 4/4 and in 5/4 time signatures correspond to different states. Likewise, the first non-downbeat in 3/4 is different from its second non-downbeat and different to any other non-downbeat in a different meter. For those~\cite{durand2017robust} who segment audio into tatum segments, time signatures of {3,4,5,6,7,8,9,10,12,16} tatums per bar are allowed. For example, considering two possible bars of two and three tatums, there would be five different states in the model. One state represents the first tatum of the two-tatum bar, and one state represents the second tatum of the two-tatum bar and so forth.

2) State transition probability matrix $\mathbf{A} = [a_{ij}]_{N \times N}$, where $a_{ij} = P(y_{t+1} = s_j | y_t = s_i), i=1,...,N; j=1, \cdots, N$ is the probability of state $s_i$ at time $t$ transferring to state $s_j$ at time $t+1$. Values of $\mathbf{A}$ needs to be trained to get (e.g. if a transition from $i$ to $j$ occurs $q$ times out of a total $Q$ transitions from $i$ to any state, then $a_{ij} = \max (\frac {q}{Q}, \quad 0.02)$).

3) Observation probability matrix $\mathbf{B} = [b_j(k)]_{N \times M}$, where $b_j(k) = P(o_t = v_k | y_t = s_j), k=1, \cdots, M; j=1, \cdots, N$ is the probability of state $s_j$ at time $t$ generates observation $v_k$. Values of $\mathbf{b}_j$ are distinguished into two cases: a) the state $s_j$ corresponds to a segment (tatum or beat) at the beginning of a bar: $s_j \in S_1 \subset S$, then it is equal to the downbeat likelihood $\mathbf{y'}$; or b) the state $s_j$ corresponds to another position inside a bar: $s_j \in \overline{S_1} \subset S$, then it is equal to the complementary probability $1 - \mathbf{y'}$:
\begin{equation}
\mathbf{b}_j =
  \left\{
   \begin{aligned}
   \mathbf{y'} & , \quad \mbox{if} \; s_j \in S_1  \\
   1 - & \mathbf{y'}, \quad \mbox{if} \; s_j \in \overline{S_1} \\
   \end{aligned}
   \right.
\end{equation}

4) Initial state probability vector ${\bm{\pi}} =[\pi _i]_{1 \times N}$, where $\pi _i = P(y_1 = s_i)$ is the probability of $y_1$ being in state $s_i$ initially. For downbeat tracking problem, each value $\pi _i$ is equally distributed: $\pi _i = \frac {1}{N}, \forall s_i \in S$.

Then we can obtain the final downbeat segment sequence following Algorithm. \ref{alg:viterbi} below.

\begin{algorithm}[htb]
\caption{Viterbi Algorithm}
\label{alg:viterbi}
\begin{algorithmic}[1] 
\REQUIRE ~~\\ 
Model $\lambda=({\mathbf{A}}, {\mathbf{B}}, {\bm{\pi}})$; observation sequence $\mathbf{o}=[o_1, o_2, \cdots, o_T]$.\\
\ENSURE ~~\\ 
Optimal state sequence $\mathbf{y} = [y_1, y_2, \cdots, y_T]$.
\label{code:init_delta}
\STATE Initialize $\delta _1(i) = {\pi _i} {b_i(o_1)}, \quad i=1, 2, \cdots, N$
\label{code:init_delta}
\STATE \qquad \qquad $\psi _1(i) = 0, \quad i=1, 2, \cdots, N$
\label{code:recursion_for}
\FOR{$t=2, 3, \cdots, T$}
\label{code:recursion_delta}
\STATE $\delta _t(i) = \max \limits_{1 \le j \le N} [\delta _{t-1}(j) a_{ji}] {b_i(o_t)}, \quad i = 1, 2, \cdots, N$
\label{code:recursion_psi}
\STATE $\psi _t(i) = \argmax \limits_{1 \le j \le N} [\delta _{t-1}(j) a_{ji}], \quad i = 1, 2, \cdots, N$
\label{code:recursion_endfor}
\ENDFOR
\label{code:termination_P}
\STATE $P^* = \max \limits_{1 \le i \le N} {\delta _T(i)}$
\label{code:termination_yt}
\STATE $y_T = \argmax \limits_{1 \le i \le N} [\delta _T(i)]$
\label{code:remount_for}
\FOR{$t=T-1, T-2, \cdots, 1$}
\label{code:remount_yt}
\STATE $y_t = \psi _{t+1} (y_{t+1})$
\label{code:remount_endfor}
\ENDFOR
\label{code:return}
\RETURN optimal sequence $\mathbf{y} = [y_1, y_2, \cdots, y_T]$; 
\end{algorithmic}
\end{algorithm}

\subsection{Dynamic Bayesian Network}\label{subsec:5.3DBN}
Dynamic Bayesian Network (DBN) is the generalization of HMM. It is adept at dealing with ambiguous RNN observations and finds the global best state sequence given these observations. DBN can use the Most Probable Explanation (MPE) feature to find the most probable state sequence. The process is analogous to the Viterbi algorithm with HMM, however, is more general. \cite{krebs2016downbeat,bock2016joint} use DBN as the temporal decoding algorithm and they model the problem as follows:

1) State space $S = \{s_1, s_2, \cdots, s_N\}$. A state $s(b, r)$ is the DBN state space is determined by two hidden state variables: the beat counter $b$ and the time signature $r$ The beat counter counts the beats within a bar $b \in \{1, \cdots ,N_r\}$ where $N_r$ is the number of beats in time signature $r$ (e.g. $r \in \{2/4, 3/4, 4/4\}$ for the case where a 3/4 and a 4/4 time signature are modelled).

2) State transition probability matrix $\mathbf{A} = [a_{ij}]_{N \times N}$. Element $a_{ij}=P(s_k | s_{k-1})$ is decomposed via:
\begin{equation}
P(s_k | s_{k-1}) = P(b_k | b_{k-1}, r_{k-1}) \dots P(r_k | r_{k-1}, b_k, b_{k-1})
\end{equation}
where
\begin{equation}
P(b_k | b_{k-1}, r_{k-1}) =
  \left\{
   \begin{aligned}
   1&, \quad \mbox{if} \; b_k=(b_{k-1} \quad mod \quad r_{k-1}) + 1  \\
   0&, \quad \mbox{otherwise} \\
   \end{aligned}
   \right.
\end{equation}
This forces that beat counter only moves steadily from left to right in a bar. Time signature changes are only allowed to happen at the beginning of a bar (i.e. $b_k < b_{k-1}$), so the probability is defined as:
\begin{equation}
\begin{aligned}
\mbox{if} \quad & b_k < b_{k-1} \\
\qquad & P(r_k | r_{k-1}, b_k, b_{k-1}) =
  \left\{
   \begin{aligned}
   1- & p_r, \quad \mbox{if} \; r_k = r_{k-1}  \\
   \frac {p_r}{R} &, \quad \mbox{if} \; r_k \neq r_{k-1} \\
   \end{aligned}
   \right. \\
\mbox{else} & \\
\qquad & P(r_k | r_{k-1}, b_k, b_{k-1}) = 0
\end{aligned}
\end{equation}
where $p_r$ is the probability of a time signature change; it is learned on the development set and \cite{krebs2016downbeat} finds out that $p_r = 10^{-7}$ is an overall good value, which makes time signature changes improbable but possible.

3) Observation probability matrix $\mathbf{B} = [b_j(k)]_{N \times M}$, where $b_j(k) = P(\mbox{features}_k | s_j)$ is the probability of state $s_j$ at time $t$ generates observation $v_k$. It can be obtained by rescale the downbeat likelihood $\mathbf{y'} = P(s_j | \mbox{features}_k)$ through:
\begin{equation}
P(\mbox{features}_k | s_j) \propto { \frac {P(s_j | \mbox{features}_k)}  {P(s_j)} }
\end{equation}

4) Initial state probability vector ${\bm{\pi}}$ is a uniform distribution over the states.

\section{Datasets}\label{sec:6_datasets}
In this section, we review the data available to downbeat tracking researches and discuss two techniques to divide datasets for training.

\subsection{Available Datasets}\label{subsec:6.1available_datasets}
Datasets\footnote{A more complete list of datasets for MIR research is at:   http://www.audiocontentanalysis.org/data-sets/} used for training and evaluation are listed in Table \ref{tab:datasets}. They are:

\begin{table*}
 \renewcommand\arraystretch{1.4}
 \caption{Overview of the available datasets for Downbeat Tracking research}
 \label{tab:datasets}
 \begin{center}
 \resizebox{\textwidth}{!}{
 \begin{tabular}{llllp{1.05\columnwidth}}
  \hline
  \textbf{Dataset}&\textbf{Reference}&\textbf{\# excerpts}&\textbf{Total length}&\textbf{Source}\\
  \hline
  Ballroom&\cite{gouyon2006experimental,krebs2013rhythmic}&685&5h 57m&http://mtg.upf.edu/ismir2004/contest/tempoContest/node5.html  https://github.com/CPJKU/BallroomAnnotations\\
  Beatles&\cite{harte2010towards,davies2009evaluation}&180&8h 09m&http://www.isophonics.net/content/reference-annotations-beatles\\
  Carnatic&\cite{srinivasamurthy2014supervised,srinivasamurthy2015particle}&176&16h 38m&http://compmusic.upf.edu/carnatic-rhythm-dataset\\
  Cretan&\cite{holzapfel2014tracking}&42&2h 20m& Not publicly available \\
  GTZAN&\cite{tzanetakis2002musical,marchand2015swing}&1000&8h 20m&http://anasynth.ircam.fr/home/media/GTZAN-rhythm/ http://www.marsyas.info/tempo/\\
  Hainsworth&\cite{hainsworth2003techniques,hainsworth2004particle}&222&3h 20m&http://www.marsyas.info/tempo/\\
  HJDB&\cite{hockman2012one}&236&3h 19m&http://ddmal.music.mcgill.ca/breakscience/dbeat\\
  Klapuri&\cite{klapuri2006analysis}&320&4h 54m&http://www.cs.tut.fi/\~klap/iiro/meter\\
  Robbie Williams&\cite{di2013automatic,giorgi2016multipath}&65&4h 31m&http://ispg.deib.polimi.it/mir-software.html\\
  Rock&\cite{de2011corpus}&200&12h 53m&http://rockcorpus.midside.com/\\
  RWC Popular&\cite{goto2002rwc,goto2004development,goto2006aist}&100&6h 47m&https://staff.aist.go.jp/m.goto/RWC-MDB/\\
  Turkish&\cite{srinivasamurthy2014search}&82&1h 33m&http://compmusic.upf.edu/corpora\\
  \hline
 \end{tabular}}
 \end{center}
\end{table*}

\textbf{Ballroom}: This dataset is (as its name implies) ballroom dancing music. It consists of 685 (after removing duplications\footnote{There are 13 duplicates which are pointed out by Bob Sturm: \url{http://media.aau.dk/null_space_pursuits/2014/01/ballroom-dataset.html}}) 30-second-length excerpts of Ballroom dance music. The total length is 5h 57m. Genres that it covers are Cha Cha, Jive, Quickstep, Rumba, Samba, Tango, Viennese Waltz, and Slow Waltz.

\textbf{Beatles}: The full name of Beatles dataset is Isophonics (Beatles only) Dataset. Songs of this dataset come from 12 studio albums of The Beatles Band. It consists of 180 excepts of the Beatles band. The total length is 8h 09m.

\textbf{Carnatic}: Carnatic dataset is short for Carnatic Music Rhythm Dataset. It is a set of art music tradition from South India. It consists of 176 songs. The total length is 16h 38m. The dataset is representative of the present day performance practice in Carnatic music and spans a wide variety of artists, forms and instruments. All labels are manually annotated. It is worth mentioning that the cultural definition of the rhythms of Carnatic music contains irregular beats.

\textbf{Cretan}: Cretan dataset is a collection of Greek music. The corpus consists of 42 full-length pieces of Cretan leaping dances. While there are several dances that differ in terms of their steps, the differences in the sound are most noticeable in the melodic content, and all pieces are considered belonging to one rhythmic style. All these dances are usually notated using a 2/4 time signature and the accompanying rhythmical patterns are usually played on a Cretan lute. While a variety of rhythmic patterns exist, they do not relate to a specific dance and can be assumed to occur in all of the 42 songs in this corpus.

\textbf{GTZAN}: GTZAN dataset was first proposed for music genre classification problem~\cite{tzanetakis2002musical}. This dataset consists of 1000 unique 30-second-length excerpts of evenly 10 genres. The total length is 8h 20m. The audio content of GTZAN dataset is representative of the real commercial music of various music genre. Also, this dataset has a good balancing between tracks with swing (blues and jazz music) and without swing.

\textbf{Hainsworth}: This dataset takes directly from CD recordings of western music. It consists of 222 excepts, and the total length is 3h 20m. Hainsworth includes six genres and styles, including choral, rock/pop, dance, classical, folk and jazz.

\textbf{HJDB}: HJDB dataset contains four genres: hardcore, jungle, and drum and bass. These are fast-paced electronic dance music genres that often employ resequenced breakbeats or drum samples from jazz and funk percussionist solos. This dataset is comprised of 236 excerpts of between 30 seconds and 2 minutes in duration. The total length is 3h 19m. Downbeat annotations were made by a professional drum and bass musician using Sonic Visualiser\footnote{http://www.sonicvisualiser.org/}.

\textbf{Klapuri}: Musical pieces of Klapuri Dataset were collected from CD recordings. Klapuri dataset consists of 320 excerpts, the total length is 4h 54m. Genres include classical, electronic/dance, hip hop/rap, jazz/blues, rock/pop, soul/R\&B/funk and unclassified. This dataset was created for the purpose of musical signal classification in general and the balance between genres is according to an informal estimate of what people listen to.

\textbf{Robbie Williams}: This dataset is composed of five albums of Robbie Williams and manual annotations. It consists of 65 songs and its total length is 4h 31m.

\textbf{Rock}: Rock dataset is based on Rolling Stone magazine's list of the "500 Greatest Songs of All Time." This dataset is still expanding with an increasing number of annotations. The newest version right now (Version 2.1) is a subset of the complete list containing 200 songs and the total length is 12h 53m.

\textbf{RWC Popular}: RWC Popular dataset with AIST Annotation is distributed as 80 Japanese popular songs with Japanese lyrics and 20 western popular songs with English lyrics. In all, this dataset consists of 100 excerpts. The total length is 6h 47m.

\textbf{Turkish}: The Turkish corpus collects Makam music from Turkey, and is an extended version of the annotated data used in \cite{srinivasamurthy2014search}. It includes 82 excerpts of one-minute length each, and each piece belongs to one of three rhythm classes that are referred to as usul in Turkish Art music. 32 pieces are in the 9/8-usul \emph{Aksak}, 20 pieces in the 10/8-usul \emph{Curcuna} and 30 samples in the 8/8-usul \emph{D\"uyek}. This dataset is composed of 230 excerpts. Turkish dataset is manually annotated. What's also worth mentioning that, the cultural definition of the rhythms contain irregular beats.

\subsection{Datasets Division Strategies}\label{subsec:6.2datasets_division}
Dataset and training technique both play a crucial role in DNN. To spy on DNN training procedure and prevent DNN from overfitting, we need to divide datasets into a training set and a development set; to check the generalization ability of DNN, we also need to divide out a test set. There are two mainstream division modes used in downbeat tracking problems: $k$-fold cross-validation and leave-one-dataset-out.

\textbf{$K$-fold cross-validation} first divides the dataset into $k$ mutually-exclusive but identically-distributed subsets of similar size. During one training procedure, $k-1$ subsets are combined as training set and the remaining one as test set; then we could obtain $k$ groups of training/test sets. After $k$ rounds of training, the mean value of $k$ results is adopted as the final result. The common $k$ value is 8.

\textbf{Leave-one-dataset-out} is recommended in \cite{livshin2003importance}, whereby in each iteration all datasets but one for training and development, and the holdout dataset for testing. After removing the test dataset, we can split 75\% for training and 25\% for development as in \cite{krebs2016downbeat}.

\section{Evaluation}\label{sec:7_evaluation}
In this section, we first describe how to evaluate whether a single segment is labeled correctly; then extend the evaluation to a whole song; finally, we take an overview of Music Information Retrieval Evaluation eXchange (MIREX) on Automatic Downbeat Estimation task and summarize the performances.

Given a predicted annotation and a known and trusted ground truth, methods of performance evaluation are required to assess algorithms and define the state of the art. The common metric is F-measure (which is also used as the evaluation method by MIREX Automatic Downbeat Estimation task), and the higher F-measure, the better model. We assume that for a specific song there exists a predicted annotation sequence $\bm{y} = \{y_1, \cdots, y_s, \cdots , y_S\}$ and a ground truth annotation sequence $\bm{g} = \{g_1, \cdots, g_t, \cdots, g_T\}$ where $S$ is the length of predicted sequence and $T$ is the length of the ground truth sequence; $S$ and $T$ may not equals. Each element value of the two sequences is time point (in seconds).

\subsection{Evaluating a Single Downbeat Label}\label{subsec:7.1eval_on_downbeat}
\begin{figure}
 \centerline{
 \includegraphics[width=6.9cm,height=2.3cm]{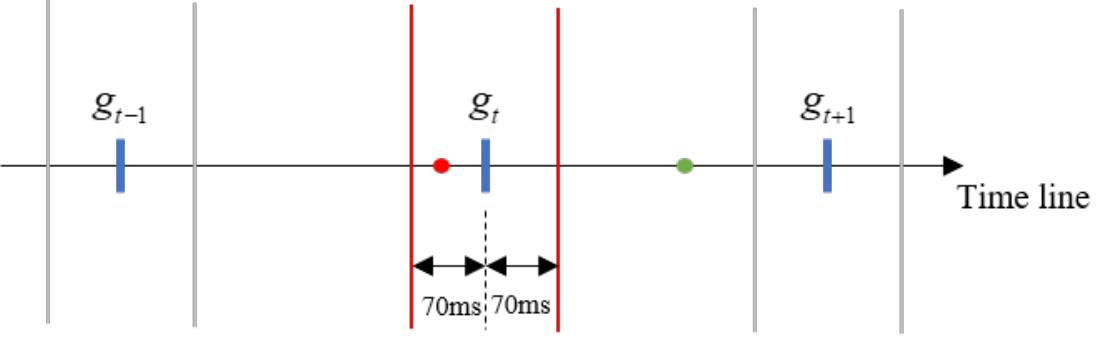}}
 \caption{An example of tolerance window (best viewed in color----\textbf{red} dot is in between $g_t$'s tolerance window while \textbf{green} dot is not).}
 \label{fig:tolerance_window}
\end{figure}

A candidate annotation $y_s$ is considered correctly tracked when it is within some fixed error window of an annotated ground truth downbeat $g_t$, where $s$ is possibly not equal to $t$, however, is the neighbor of $t$. This window is called the \emph{tolerance window}, and the common size is $\pm$70 ms. For instance, in Fig. \ref{fig:tolerance_window}, if a predicted downbeat $y_s$ meets certain $g_t$'s tolerance window: $({g_t - 70ms}) \le y_s \le ({g_t + 70ms})$ (located between two vertical red lines) , it is a \emph{true positive} (just like the red dot).

\subsection{Evaluating on a Song}\label{subsec:7.2eval_on_song}
A predicted annotation is perfectly correct if it is a true positive. If a predicted annotation is not in the tolerance window of any ground truth annotation, it is a false positive (just like the green dot in Fig. \ref{fig:tolerance_window}). The number of false negatives is counted in a tricky way: if a ground truth annotation has no predicted annotations meeting its tolerance window (just like $g_{t-1}$ and $g_{t+1}$ in Fig. \ref{fig:tolerance_window}), the amount of false negatives increases by one. Obviously, there is a vacant predicted annotation in its tolerance window, which is supposed to be a false negative.

Add up all statistics of a specific song by comparing $\bm{y}$ and $\bm{g}$. The number of true positives $tp$, false positives $fp$ and false negatives $fn$ are combined to calculate \emph{precision} and \emph{recall}:
\begin{equation}
precision = \frac { tp }{ tp + fp }
\end{equation}
\begin{equation}
recall = \frac {tp}{tp + fn}
\end{equation}
Then the F-measure on a song is computed as:
\begin{equation}
F measure = \frac {2 \times precision \times recall}{precision + recall}
\end{equation}

Some researches~\cite{durand2017robust} don't take into account the first 5 seconds and the last 3 seconds of audio when evaluating a song. Because annotations are sometimes missing or not always reliable.

\subsection{Music Information Retrieval Evaluation eXchange}\label{subsec:7.3mirex}
Since 2014, downbeat estimation systems have been compared in an annual evaluation held in conjunction with the \emph{International Society for Music Information Retrieval}\footnote{ \url{http://www.music-ir.org/mirex/wiki/MIREX\_HOME} }. Authors submit algorithms which are tested on several datasets of audio and ground truth. For downbeat estimation systems that require training, the dataset is split into a training set for training and a test set for evaluating the performance. We present a summary of the algorithms submitted in Table \ref{tab:mirex_systems}. Due to the high diversity of musical styles among these datasets, performances of all algorithms are reported per each individual dataset. Note that results of the year 2017 and 2018 haven't come out.

\begin{table*}
 \renewcommand\arraystretch{1.4}
 \begin{minipage}{\textwidth}  
 \caption{MIREX Systems from 2014-2017, sorted in each year by F-measure evaluation. The best system in each dataset in that year are \underline{underlined}. The best result in each dataset over the years are shown  in \textbf{bold font}. Systems where no data is available are shown by a dash (-). Results marked by an asterisk should be taken with care as in those cases overlapping test and training sets were used.}
 \label{tab:mirex_systems}
 \begin{center}
 \resizebox{\textwidth}{!}{
 \begin{tabular}{lccp{0.36\columnwidth}cccccccc}
  \hline
  \multirow{2}*{\textbf{Year}}&\multirow{2}*{\textbf{Submission Code}}&\multirow{2}*{\textbf{Abstract}}&\multirow{2}*{\textbf{Approach(es)\footnote{If two approaches are listed, the first one represents DNN-based feature learning algorithm and the second one refers to the temporal decoding method. Note that not all of the MIREX systems are deep learning-based, we still list here. And if just one approach is listed, it denotes the temporal decoding method.}}}&\multicolumn{8}{ c }{\textbf{Performance (F-measure)}}\\
  \cline{5-12}
  ~&~&~&~&\textbf{Ballroom}&\textbf{Beatles}&\textbf{Carnatic}&\textbf{Turkish}&\textbf{Cretan}&\textbf{HJDB}&\textbf{RWC\_classical}&\textbf{GTZAN}\\
  \hline
  2014&DBDR2&\cite{MIREX2014DBDR}&deep belief network; Viterbi algorithm&0.705&\underline{0.831}*&0.184&0.448&0.435&0.435&-&-\\
  ~&DBDR3&\cite{MIREX2014DBDR}&deep belief network; Viterbi algorithm&0.752*&0.816&0.2&0.448&0.415&0.415&-&-\\
  ~&FK3&\cite{MIREX2014FK3}&dynamic Bayesian network&\underline{0.792}*&0.588&0.169&0.197&0.535&0.535&-&-\\
  ~&FK4&\cite{MIREX2014FK4}&dynamic Bayesian network&0.708*&0.63&0.194&0.24&0.512&0.512&-&-\\
  ~&KSH1&\cite{MIREX2014KSH1}&hidden Markov model&0.194&0.194&\underline{0.4}&\underline{\textbf{0.775}}*&\underline{\textbf{0.854}}*&\underline{0.854}&-&-\\
  \hline
  2015&DBDR2&\cite{MIREX2015DBDR}&CNN; Viterbi algorithm &0.763&\underline{0.855}*&0.221&\underline{0.472}&0.415&0.691&-&-\\
  ~&DBDR3&\cite{MIREX2015DBDR}&CNN; Viterbi algorithm &\underline{0.802}*&0.847&0.216&0.446&0.449&0.682&-&-\\
  ~&FK2&\cite{MIREX2015FK}&hidden Markov model &0.503&0.713*&0.154&0.289&0.151&0.794&-&-\\
  ~&FK3&\cite{MIREX2015FK}&hidden Markov model &0.595*&0.709&0.166&0.298&0.167&\underline{0.824}&-&-\\
  ~&FK4&\cite{MIREX2015FK}&hidden Markov model &0.179&0.178&\underline{\textbf{0.474}}&0.142&0.233&0.12&-&-\\
  ~&FK6&\cite{MIREX2015FK}&hidden Markov model &0.756*&0.642&0.197&0.284&\underline{0.529}&0.626&-&-\\
  \hline
  2016&DBDR1&\cite{MIREX2016DBDR1}&CNN; Viterbi algorithm &0.838*&0.849&0.201&0.306&0.426&0.578&0.527*&0.615\\
  ~&DBDR2&\cite{MIREX2016DBDR1}&CNN; Viterbi algorithm &0.783&\underline{\textbf{0.872}}*&0.231&0.415&0.418&0.629&0.532*&0.619\\
  ~&KB1&\cite{MIREX2016KB}&RNN; hidden Markov model  &0.898*&0.803&0.269&0.352&0.433&0.69&0.436&0.63\\
  ~&KB2&\cite{MIREX2016KB}&RNN; hidden Markov model &0.86*&0.818*&0.33*&0.336*&0.443*&0.851*&0.428*&\underline{\textbf{0.647}}\\
  ~&BK4&\cite{MIREX2016BK4}&RNN; dynamic Bayesian network &\underline{\textbf{0.908}}*&0.865*&\underline{0.369}*&\underline{0.537}*&\underline{0.635}*&\underline{\textbf{0.97}}*&\underline{\textbf{0.599}}*&0.638\\
  ~&DSR1&\cite{MIREX2016DSR1}&Viterbi algorithm &0.463&0.665&0.184&0.317&0.265&0.208&0.251&0.397\\
  ~&CD4&\cite{MIREX2016CD4}&Viterbi algorithm&0.412&0.604&0.186&0.218&0.25&0.334&0.174&0.46\\
  \hline
 \end{tabular}}
 \end{center}
 \end{minipage}  
\end{table*}

\subsubsection{MIREX 2014}
In Audio Downbeat Estimation task of MIREX 2014, six datasets were used to train and test the submitted algorithms. Audio in these datasets is monophonic sound files of CD-quality (PCM, 16 bit, 44100 Hz) except Ballroom (originally lower quality, but resampled to 44100 Hz).

Krebs's submission FK3 achieved an F-measure of 0.792 on Ballroom dataset by using Dynamic Bayesian Network. Durand et al.'s submission DBDR2 achieved 0.831 on Beatles dataset, using Deep Belief Network and Viterbi Algorithm. Submission KSH1 of Krebs, \\* Holzapfel and Srinivasamurthy obtained the highest performance on four datasets: Carnatic, Turkish, Cretan and HJDB, with F-measure of 0.4, 0.775, 0.854 and 0.854 respectively. Algorithms used in KSH1 are bar pointer model~\cite{holzapfel2014tracking} and HMM.

\subsubsection{MIREX 2015}
Datasets used in 2015 was as same as last year. Durand et al.'s submission DBDR3 reached 0.802 on Ballroom dataset and submission DBDR2 0.855 on Beatles dataset using DNNs and Viterbi algorithm. Krebs and B\"ock's submission FK3 obtained an F-measure of 0.824 on HJDB dataset by using HMM. From an overall perspective, most submissions performed better on Ballroom, Beatles and HJDB datasets. Audio in these datasets basically is western music; while those in the other three datasets are non-western music--whose time signature and tempo range are not quite regular.

\subsubsection{MIREX 2016}
In 2016, the number of datasets is increased to eight, and the new datasets are RWC classical and GTZAN. By now, performance had steadily risen from early work in 2014. The first thing to notice from Table \ref{tab:mirex_systems} is that performances on two datasets had reached above 0.9: submission BK4 of B\"ock and Krebs obtained 0.908 on Ballroom and 0.97 on HJDB by using RNN and Dynamic Bayesian Network. Durand et al.'s submission DBDR2 achieved 0.872 on Beatles dataset using DNN and Viterbi algorithm. Performances on the two new datasets are not perfectly well, with F-measure of 0.599 (submission BK4) on RWC classical and 0.647 (submission KB2) on GTZAN.

\subsection{Summary and Evolution of MIREX Performance}\label{subsec:7.4summary_mirex}
\begin{figure}
 \centerline{
 \includegraphics[width=\columnwidth,height=5cm]{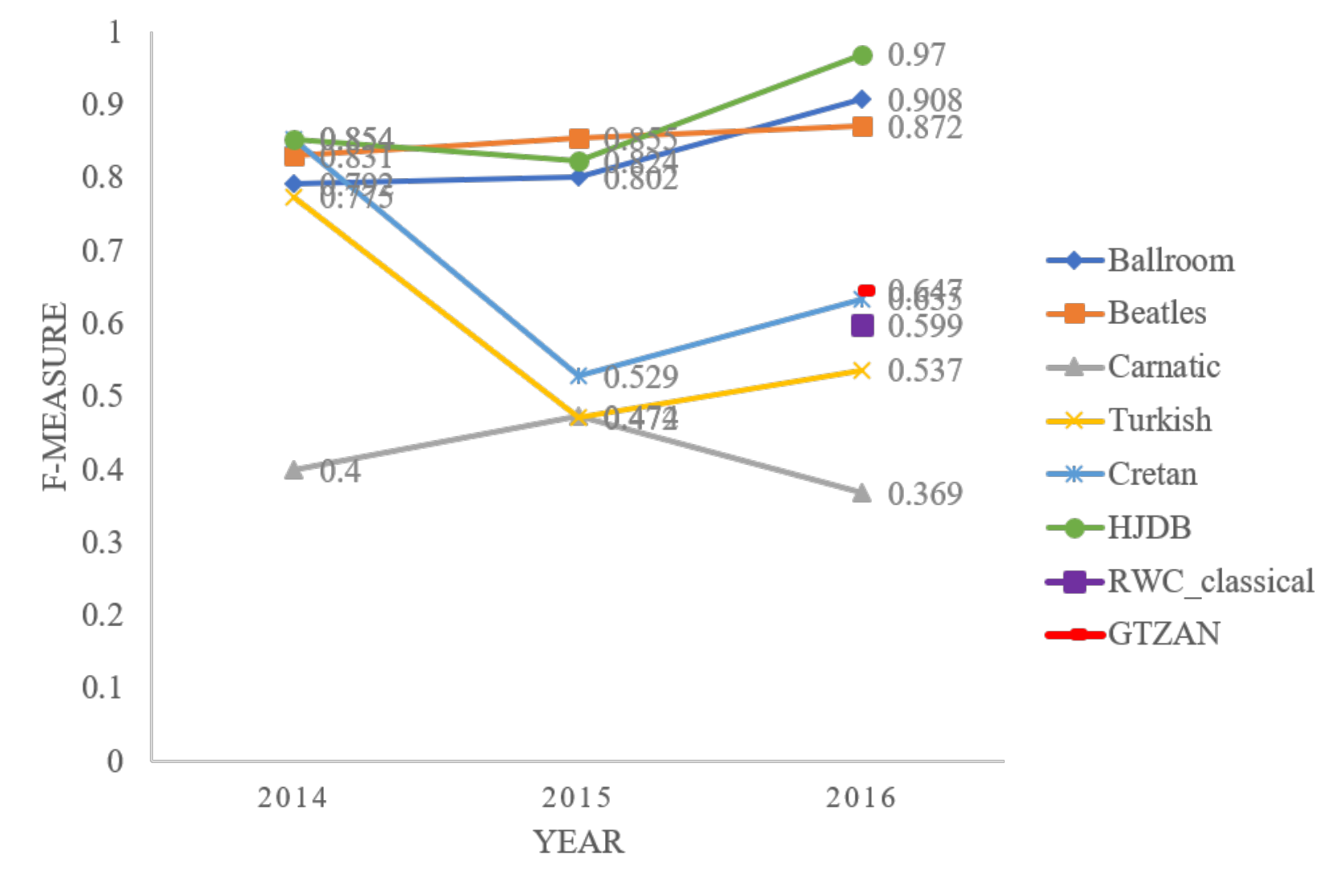}}
 \caption{The evolution of the best performance (F-measure) per dataset per year on Automatic Downbeat Estimation task of MIREX (best viewed in color).}
 \label{fig:mirex_best_performance}
\end{figure}

We show the annual evolution of the best performance on each dataset of the MIREX Automatic Downbeat Estimation task as a line chart displayed in Fig. \ref{fig:mirex_best_performance}. We can see from this figure that the performances on three western music datasets (Ballroom, Beatles and HJDB) have slightly increased from the year 2014 to 2016; while the F-measures on non-western music datasets (Carnatic, Cretan, Turkish) have declined as a whole\footnote{Here we only talk about the six datasets used since 2014 because there are no comparisons for RWC classical and GTZAN}. It seems that algorithms usually fail on non-western music, and the reason could be two-fold: a) Comparing to the number of western excerpts (1101 tracks), the number of non-western excerpts (300 tracks) is few, which leads to an imbalanced training set.  b) Time signatures of most tracks in three western music are commonly used (2/4, 3/4 and 4/4), whereas the time signatures in the other two non-western music are various and rare (Carnatic art music contains 5/4 and 7/4 meters; Turkish contains 8/8, 9/8 and 10/8 usul).  What's worth mentioning that music in Cretan dataset truly is 2/4 time signature but the volume of this dataset is too small (40 tracks), therefore learning algorithms could hardly learn features.

\section{Software Packages}\label{sec:8_software}
A few software packages or toolboxes are released over the years to solve downbeat tracking problems. In this section, we summarize an incomplete list of the most relevant packages.

Madmom\footnote{\url{https://github.com/CPJKU/madmom}}, first released in 2016, is an open-source audio signal processing library written in Python with a strong focus on MIR tasks~\cite{bock2016madmom}. Apart from focusing on low-level music features, madmom puts emphasis on musically meaningful high-level features by implementing some signal processing methods. Also, madmom provides a module that implements some in MIR commonly used machine learning methods such as HMM and DNN; and it comes with several state-of-the-art MIR algorithms for onset detection, beat, downbeat and meter tracking, tempo estimation, and piano transcription.

There are other toolboxes or packages that are quite relevant to downbeat tracking. MIRtoolbox\footnote{\url{https://www.jyu.fi/hytk/fi/laitokset/mutku/en/research/materials/mirtoolbox}} is a free (to the research community) Matlab toolbox dedicated to the extraction of musically-related features from audio recordings such as tonality, rhythm, structures, etc. Additionally to some basic computational approaches for low- and mid-level features, the toolbox also includes higher-level musical feature extraction tools~\cite{lartillot2007matlab,latrillot2007mir}. Essentia\footnote{\url{http://essentia.upf.edu/documentation/}} is an open-source C++ library (also wrapped in Python) for audio analysis and audio-based music information retrieval~\cite{bogdanov2013essentia}. It contains an extensive collection of reusable algorithms which implement audio input/output functionality, standard digital signal processing blocks, statistical characterization of data, and a large set of spectral, temporal, tonal and high-level music descriptors. LibROSA\footnote{\url{https://github.com/librosa/librosa}} is a python package for audio and music signal processing; it provides the building blocks necessary to create music information retrieval systems~\cite{mcfee2015librosa}. It covers core input/output audio processing and digital signal processing functions, visualization, structural segmentation, feature extraction, and manipulation etc.

\section{Discussion}\label{sec:9_discussion}
As a continuous research area, automatic downbeat tracking has received quite an amount of attention in academic researches and for industrial applications. It is aimed to annotate all downbeat time points in the music, so that users can precisely follow the groove while listening to it or can easily divide a music piece into bars etc. A concise chronological review of the associated literature in DNN-based downbeat tracking, together with the main contributions of each work, according to the timeline, is shown in Table \ref{tab:chronological_list}.

\begin{table*}
 \renewcommand\arraystretch{1.4}
 \caption{Chronological Summary of Advances in DNN-Based Downbeat Tracking, Years 2015-2017, Showing Year of Publication, Reference Number, Authors, Title and (step-by-step) Methods to the Field.}
 \label{tab:chronological_list}
 \begin{center}
 \resizebox{\textwidth}{!}{
 \begin{tabular}{lp{0.16\columnwidth}lp{0.7\columnwidth}p{0.7\columnwidth}}
  \hline
  \textbf{Year}&\textbf{Reference Number}&\textbf{Authors}&\textbf{Title}&\textbf{Methods}\\
  \hline
  2015&\cite{durand2015downbeat}&S. Durand, et al.&Downbeat Tracking With multiple Features and Deep Neural Networks&beat segmentation; multiple features extraction; DNNs; Viterbi Algorithm\\
  \hline
  2016&\cite{bock2016joint}&S. B\"ock, et al.&Joint Beat and Downbeat Tracking with Recurrent Neural Networks&frame segmentation; auto-learned features; RNN; DBN\\
  &\cite{krebs2016downbeat}&F. Krebs, et al.&Downbeat Tracking Using Beat-Synchronous Features and Recurrent Neural Networks&beat segmentation; percussive and harmonic features; RNNs; DBN\\
  &\cite{durand2016feature}&S. Durand, et al.&Feature Adapted Convolutional Neural Networks for Downbeat Tracking&tatum segmentation; rhythm, melodic and harmony feature extraction; CNNs; HMM\\
  \hline
  2017&\cite{durand2017robust}&S.Durand, et al.&Robust Downbeat Tracking Using an Ensemble of Convolutional Networks&tatum segmentation; multiple features extraction; CNNs; Viterbi Algorithm\\
  \hline
 \end{tabular}}
 \end{center}
\end{table*}

\subsection{Detailed Analysis}\label{subsec:9.1analysis}
An analysis of each key step of a prevalent system is stated below.

\subsubsection{Segmentation}
Beat segmentation is the always first thought because normally the first beat of a bar is downbeat. One can easily think of the way to find downbeat by deciding a beat is a downbeat or not. However, automatic beat tracking is still not perfect, even though \cite{durand2015downbeat} tries to ease this problem by way of seeking the segmentation that maximizes downbeat recall while emphasizing consistency in inter-segmentation durations.

Tatum is a more fine-grained temporal unit. There are three reasons using tatums: a) tatum encodes a musically meaningful dimension reduction according to tempo invariance, b) tatum reduce the cost of designing, training and testing DNN and temporal decoding algorithms, and c) comparing to beat segmentation, tatum segmentation achieves higher recall rate, enabling almost all possible downbeats under detection. However, Durand et al. also point out in \cite{durand2017robust} that tatum segmentation has a downbeat recall rate of 92.9\% considering a $\pm$70 ms tolerance window and therefore occasionally misses an annotated downbeat. Another problem in tatum segmentation pointed by \cite{durand2017robust} is that two consecutive bars may contain a different number of estimated tatums.

A frame is just a raw segment of the original audio. It is not a temporal unit in the metrical level of music. Nevertheless, comparing to tatums, segmenting audio into frames takes every piece of music as a downbeat candidate and indeed won't miss a downbeat. But it obviously increase the number of samples. So frame segmentation needs to cooperating with automatic feature extraction method and better works with DNN.

\subsubsection{Features Selection}
The effectiveness of the feature extraction part depends on the selection of features. Which feature is actually contributing is not very clear. Features mentioned in section \ref{sec:3_feature_extraction} are mostly hand-crafted and are considered to be related to downbeats in experts' view. We don't analysis the feature extraction methods here since they are common methods in audio signal processing for music applications. We will discuss the effect of feature design at the general level.

Durand et al.~\cite{durand2015downbeat} have done a series of ablation studies to testify the importance of features. In their experiments, they ran a simplified version of the system without temporal decoding step. They added one feature at a time while conducting each experiment (the order of features added is not important). The F-measure result increases as they add features and adding all features increases 18 F-measure scores comparing to average. The result of this study adheres to our intuition since each possibly downbeat-related feature contributes a little to the final performance. Nonetheless, every automatic downbeat tracking system chooses different features to use, basically according to the researchers' intuition.

Automatic learned feature exceeds hand-crafted feature because it doesn't rely on human intuitions and doesn't exist human prejudice. Relevant features are directly learned from the raw audio signal by using a feature learning algorithm. In this setting, a good feature learning algorithm is particularly important. The quality of the model directly influences the selection of features and further impacts on the final performance.

\subsubsection{DNN-based Feature Learning}
To testify whether DNN-based feature learning method is necessary, some researchers have also conducted several ablation experiments~ \cite{durand2015downbeat,durand2017robust}. In \cite{durand2017robust}, researchers compare the deep learning method with a shallow learning method SVM and results show an improvement of around 10 points of F-measure. In \cite{durand2015downbeat}, researchers fix all the features and the temporal decoding step, comparing the feature learning method between the DNN and a linear regression method. Their results show that there is a 12-point increase in the F-measure score when using DNN, which is statistically significant. The system in \cite{durand2015downbeat} is compared to three non-DNN downbeat tracking systems~\cite{davies2006spectral,peeters2011simultaneous,papadopoulos2011joint}. Their system achieves a mean F-measure of 67.5 points compared to other three non-DNN systems (48.7 points in \cite{peeters2011simultaneous}, 51.7 points in \cite{davies2006spectral}, 52.2 points in \cite{papadopoulos2011joint}). Taken dataset individually, DNN-based system doesn't improve much (about 10 points) when the dataset is relatively small since in this case, a simple learning algorithm can already give good results. However when the dataset is more complex (fewer clues, more changes in time signature, soft onsets or where there is not always percussion), the DNN-based system improves a lot (about 19 points). Note that these systems all fail in certain datasets where there are expressive timings because bar boundaries are not clear and distinguishable.

Results shown in Table \ref{tab:mirex_systems} give a clear comparison between several prevalent systems\footnote{Note that almost all researchers of automatic downbeat tracking have participated in MIREX Automatic Downbeat Estimation task and systems they proposed in their papers are similar to ones in MIREX. So results in Table \ref{tab:mirex_systems} are quite representative and sufficient enough to analysis}. We can see a trend of using DNN-based feature learning algorithm through years and also see F-measure scores increase on the whole through years. To make an unambiguous analysis, the comparison is made among systems focusing on different datasets. For Ballroom, Beatles, HJDB and GTZAN datasets, results achieves a relatively high F-measure score when using DNN-based learning methods comparing to shallow methods, all surpassing 0.6 points and some even reaching above 0.9 points. This is because these datasets are of large data size, small variance, common time signatures, hard onsets, and distinct percussions. For other datasets which do not possess the above attributes, like Carnatic (containing irregular beats), Turkish (unusual time signatures), Cretan (small data size) and RWC\_classical (soft onsets and blurry percussions) datasets, DNN-based systems performance a little worse than shallow ones, however generally all these systems performance not very well. In summary, DNN exceeds other learning algorithms in the aspect of learning high-level feature representations in a data-driven circumstance. Comparing to deep models, shallow ones are less able to classify segment with the perceptively correct results moving towards out-of-phase or inconsistent segments.

\subsubsection{Temporal Decoding}
The temporal model plays an important role of further boosting the performance of DNN. To testify this, Durand et al.~\cite{durand2017robust} conduct a comparison study where they remove the temporal decoding step with a hard threshold. In the configuration without temporal decoding, a position is a downbeat if its likelihood exceeds a fixed oracle threshold. The threshold $t=0.88$ is manually set to achieve the best F-measure and it corresponds roughly to the ratio of downbeats and non-downbeats in the dataset. Results show the system with temporal decoding surpasses over 10 points than that with the threshold. This can be interpreted as the raw output of DNN is a noisy downbeat likelihood sequence.

\subsection{Future Work}\label{subsec:9.2future_work}
Despite the success of DNN-based Downbeat Tracking Systems and considerable effort that many researchers have made, many problems still need to be addressed in automatic downbeat tracking before these techniques can be applied to a wide range of complex real-world problems. Problems that need to be solved are: the relatively lower results for classical music dataset and for songs where there are expressive timings (time signature changes within a musical piece)~\cite{durand2015downbeat, bock2016joint}, the lack of the diversity of time signatures in the used datasets~ \cite{durand2017robust} (some even need to know the time signature in advance~\cite{krebs2016downbeat}), the uncertainty of effectiveness of manually selected features. This section summarizes these issues and accordingly discusses future research direction.

\subsubsection{Improving datset quality}\label{subsec:9.2.1}
DNN-based models are very limited to the integrity, variety, richness, exhaustiveness, and balance of training datasets. They will perform better for the sake of better datasets. Therefore, the quality of datasets is extremely important, especially the size, diversity, and balancing of datasets matter the most. However, none of the existing datasets has satisfied this requirement.

First of all, the magnitude and size of existing datasets is so small (for example Cretan and Robbie Williams dataset only consists of 42 and 65 songs respectively) that the information provided for deep learning method is not enough. Second and third, the lack of diversity (especially of time signatures) and balancing is also a severe issue. Among the available datasets, western music is in the majority, and under most circumstances, the time signatures used in western music are 3/4 and 4/4. Even though there are Indian (Carnatic dataset), Greek (Cretan dataset) and Turkish (Turkish dataset) music, the time signatures are pretty rare or little (Carnatic: 5/4 and 7/4 meters; Greek: 2/4 time signature; Turkish: 8/8, 9/8 and 10/8 usul). These issues are quite obviously revealed in Fig. \ref{fig:mirex_best_performance} since we can see that the performances on western music dataset are better than non-western music datasets as a whole. Since the downbeat position is highly relevant to time signature, datasets with unbalanced time signatures will significantly hinder deep learning methods performance. Last but not least, the variety and richness of the available datasets are not wide enough. For songs of different genres and various forms of expression, their downbeat traits are also very different. When facing more complex datasets, where there are fewer clues, more changes in time signature, soft onsets or where there is not always percussion, such as Classical, Jazz or Klapuri subset datasets, the results are relatively lower~ \cite{durand2015downbeat}. In regard to the limitation of the system of not being able to perform time signature changes within a musical piece, particle filters as used in \cite{krebs2015inferring} should be able to solve this problem~\cite{bock2016joint}.

There is another issue that needs pointing out. As mentioned in Section \ref{subsec:6.2datasets_division} before, dataset division strategy is crucial to training procedure of DNN, and researches in automatic downbeat tracking haven't used the same division strategy, which will make the performance comparison less convincing. Therefore, defining standardized dataset train/test split is also an urgent task. Future work should refine and organize more and better datasets, in terms of the size, diversity, balancing and standardized split of datasets. Albeit, dataset labeling, and organization is both labor-consuming and time-consuming, more and more contributions are still needed.

\subsubsection{Data augmentation}\label{subsec:9.2.2}
Another way to solve dataset problem is to do data augmentation. This could be faster than the solution of improving dataset quality. Data augmentation has been widely used in deep learning tasks because one of the essential requirements of deep learning is a huge amount of data. When the dataset is inadequate and unbalanced, data augmentation can be a good approach to increase the size of data. Data augmentation can also increase the diversity of dataset to prevent the model from overfitting (simply memorizing music sequence~\cite{mor2018universal}). For music audio, possible data augment strategies could be pitch shifting~\cite{mor2018universal}, time-scale modification. As long as the innate downbeat characteristics stay unchanging, we can do any augmentation to widen dataset scale.

\subsubsection{Automatic feature discovery}\label{subsec:9.2.3}
Hand-crafted features are extracted according to human's domain knowledge. However, these features have not proven to be highly correlated to downbeat and their effectiveness and validity are not very clear. In terms of the definition of downbeat, which is the first beat of each bar, we speculate that downbeat is in high correlation with time. More specifically, attributes related to the bar are possibly related to downbeat as well, such as \emph{tempo} and \emph{time signature}. If we know the music audio duration (time length), tempo, time signature and time stamp of the first downbeat, we could reckon all downbeat positions in this audio (assume that there are no rhythm flexibility because it will cause inequality of each bar). Nevertheless, these attributes are also unknown in advance, let alone there could exist rhythm flexibility.

Straightforwardly, we can calculate tempo and time signature first, then use them to calculate the downbeat position or guide learning algorithms as conditions. However, this approach relies on the precision and accuracy of the estimated tempo and time signature, otherwise, errors will be introduced. Another approach is automatically learning features, which~\cite{bock2016joint} has already tried to use. But \cite{bock2016joint} still applies some human's prior knowledge as they preprocess audio with specified hand-made digital signal processing procedure. To achieve complete automatic feature discovery and extract attributes from scratch without any human guidance, we can use a novel deep learning architecture to learn attributes all by itself~\cite{zhang2014start}, to mine useful higher-level representations and use them as inputs to feed learning model.

\subsubsection{Improving deep learning architecture}\label{subsec:9.2.4}
From another perspective, we can see that models used in downbeat tracking system are not powerful enough. Since the researches of deep learning have exploded, more advanced models appear. On one hand, we can focus on replacing the basic DNN models in the system of more advanced DNN models. Possible effective models include dilated CNN (which excels at extracting features in a wider-range), dilated RNN (which is good at modeling both short-term and long-term time series) and highway networks etc. In time, we also hope that our theoretical understanding of the properties of neural networks will improve, as it currently lags far behind the practice. On the other hand, a network combination procedure adapted to the temporal model seems promising to improve performance~\cite{durand2016feature, durand2017robust}. Moreover, downbeats of some songs are not quite related to the aforementioned hand-crafted features, then maybe we could combine feature extraction and feature learning parts and let deep learning algorithms process together. And this leads to a more adventurous way--using end-to-end neural network to merge all stages together and process the whole system by only designing a powerful neural network architecture. End-to-end deep architectures~\cite{dieleman2014end, miao2015eesen, ma2016end, zhang2016play} are feasible and alternative approaches to combine these two stages (feature extraction and feature learning). As a general rule, features are extracted from music audio signals and are then used as input to a learner, such as deep neural networks. The features are designed to uncover information in the input that is salient for the task at hand. This requires considerable expertise about the problem and constitutes a significant engineering effort. In this case, end-to-end models require no feature engineering or complex data preprocessing, thus making it applicable to automatic downbeat tracking problem. Using end-to-end architecture covers the solution to the problem described in section \ref{subsec:9.2.3} as it obviously combines that part of the architecture.

\section{Conclusion}\label{sec:10_conclusion}
Automatic downbeat tracking is to find out the temporal locations of all downbeats in music audio. It is a promising task for the sake of the music industry, musicians and music lovers, and for them to better understand, process and learn music. Enabling machines to possess the capability of perceiving music is a difficult task. Hence, researchers are attempting to establish an automatic downbeat tracking system using various methods. To conclude, it is worth revisiting the overarching goal of all of this research: reviewing the current automatic downbeat tracking systems based on several kinds of deep neural networks, mostly DNN, CNN, and RNN. We detail every procedure of downbeat tracking system step by step in this work. To start, we describe the preprocessing phrases, including all the segmentation methods and all the features extracted from music data. Next, we depict every deep neural network used in the feature learning part, both visually and theoretically. Subsequently, temporal decoding methods used at the end of the system are summarized. In addition, to provide researchers with an easy way to use the public downbeat dataset, we collect and organize all the information of the available datasets in this task. Furthermore, standardized and acknowledged evaluation metrics used in automatic downbeat tracking are described. We also discussed some available software and APIs. Finally, we summarize and point out some existing problems in current researches, and put forward some suggestions and possible solutions for future research directions.

\begin{acknowledgements}
This work is supported by National Natural Science Fund for Distinguished Young Scholar (Grant No. 61625204) and partially supported by the State Key Program of National Science Foundation of China (Grant Nos. 61836006 and 61432014).
\end{acknowledgements}



\end{document}